\documentclass[aps,prb,twocolumn,showpacs]{revtex4}
\usepackage{ifthen}
\newcommand{\usepdfs}{false}
\ifthenelse{\equal{\usepdfs}{true}}{
  \usepackage[pdftex]{graphicx}
}
{
  \usepackage[dvips]{graphicx}
}
\ifthenelse{\equal{\usepdfs}{true}}{
  \newcommand{\ifig}[2][]{\includegraphics[#1]{#2.pdf}}
}
{
  \newcommand{\ifig}[2][]{\includegraphics[#1]{#2}}
}
\usepackage{amssymb,amsmath,amsfonts,hyperref}
\usepackage{wasysym}
\usepackage{latexsym}
\usepackage[dvips]{graphicx}

\newcommand{\ket}[1]{| #1 \rangle}
\newcommand{\bra}[1]{\langle #1 |}

\newcommand{\I}{{\hat {\mathcal R}}_{(2)}}
\renewcommand{\P}{{\hat {\mathcal R}}_{(3)}}
\newcommand{\C}{{\hat {\mathcal P}}}
\newcommand{\T}[1]{{\hat T}_{#1}}
\newcommand{\R}{{\hat {\mathcal I}}}

\newcommand{\A}{\mathcal A}
\renewcommand{\u}{u}
\renewcommand{\d}{d}

\newcommand{\divo}{\operatorname{div}}
\newcommand{\curl}{\operatorname{curl}}

\newcommand{\be}{\begin{equation}}
\newcommand{\ee}{\end{equation}}
\newcommand{\bs}{\begin{split}}
\newcommand{\spl}{\end{split}}

\newcommand{\cdaf}{CdCr$_2$O$_4$}
\newcommand{\hgaf}{HgCr$_2$O$_4$}

\begin{document}

\title{Ordering in a frustrated pyrochlore antiferromagnet proximate to a spin liquid} \date{\today}

\author{Doron L. Bergman,$^1$ Gregory A. Fiete,$^2$ and Leon Balents$^1$}
\affiliation{$^1$Department of Physics, University of California, Santa Barbara, CA
93106-9530\\$^2$Kavli Institute for Theoretical Physics, University of 
California, Santa Barbara, CA 93106-4030}

\begin{abstract}
  We perform a general study of spin ordering on the pyrochlore
  lattice with a 3:1 proportionality of two spin polarizations.
  Equivalently, this describes valence bond solid conformations of a
  quantum dimer model on the diamond lattice.  We determine the set of
  likely low temperature ordered phases, on the assumption that the
  ordering is weak, i.e the system is close to a ``U(1)'' quantum spin
  liquid in which the 3:1 proportionality is maintained but the
  spins are strongly fluctuating.  The nature of the 9 ordered states
  we find is determined by a ``projective symmetry'' analysis.  All
  the phases exhibit translational and rotational symmetry breaking,
  with an enlarged unit cell containing 4 to 64 primitive cells of the
  underlying pyrochlore.  The simplest of the 9 phases is the same
  ``R'' state found earlier in a theoretical study of the ordering on
  the magnetization plateau in the $S=3/2$ materials \cdaf~and \hgaf.
  We suggest that the spin/dimer model proposed therein undergoes a direct
  transition from the spin liquid to the R state, and describe a field theory for the
  universal properties of this critical point, at zero and non-zero
  temperatures.  
\end{abstract}
\date{\today}
\pacs{64.60.-i,71.10.-w,71.27.+a}

%64.60.-i General studies of phase transitions
%64.60.Cn order-disorder transformations; statistical mechanics of model systems
%71.10.-w Theories and models of many-electron systems
%71.27.+a Strongly correlated electron systems; heavy fermions

%\email{doronber@physics.ucsb.edu}
%\keywords{Fractionalization; CDW; Duality; pyrochlore lattice}

\maketitle

%\tableofcontents
%\numberwithin{equation}{subsection}

\section{Introduction}

Charge and/or magnetic order is an apparently central feature of the
ground states of Mott insulators.  The detailed nature of this order
can be difficult to understand theoretically, particularly when
``frustration'' is present.  By frustration, we mean the 
presence of competing interactions, which lead, in some appropriate
``classical'' limit, to a large number of degenerate ground states.
This degeneracy is lifted by fluctuations, thermal and quantum, or
additional interactions beyond those of the classical limit.  However,
the classical degeneracy can be lifted in many different ways, making
the ultimate ground state very sensitive to details of the
Hamiltonian.  Apart from demanding an extremely detailed microscopic
understanding of a given material (always desirable, but not so easy
to come by!), is there any way to attack such problems?  The approach
we follow in this paper is to presume that fluctuations (in this case
quantum) are strong, which requires that the ordering itself is weak
(i.e. the charge and/or spin modulations are small in amplitude).  If
so, we may presume the system to be close to some ``liquid'' state, in
which no order (in the conventional sense -- but see below) is
present.  One may then explore the possible ordered states which
occur as weak instabilities of the liquid.

Recently, such a view to {\sl charge} ordered states of {\sl
  two-dimensional} lattice {\sl boson} systems has been systematically
pursued in Refs.~\onlinecite{Balents_1:prb05,Balents_2:prb05}, with
specific precedents in Refs.~\onlinecite{Lannert:prb01,Sachdev:ann02}.
In that case, the liquid state was taken to be a superfluid.  There, the
possible charge ordered Mott insulating states proximate to the
superfluid were discussed by considering the instabilities due to
proliferation of {\sl vortices}.  This was made systematic by uncovering
the multiplet structure of the vortex states, determined by symmetry.
In particular, vortices were shown to transform under a {\sl projective
  representation} of the lattice space group, or Projective
Space/Symmetry Group (PSG)\footnote{ The PSG is useful as a means 
of classifying emergent gauge structures as discussed by X. G. Wen in Ref.~\onlinecite{Wen:prb02}
}.  The PSG was shown to determine the
structure of the action of the critical theory for the superfluid-Mott
transition, and the nature of the possible charge ordered Mott phases.
The PSG depends upon the lattice symmetries (space group) {\sl and} the
mean conserved boson density.  All the above
considerations for bosons apply equally well to spin models with
$U(1)$ rather than full $SU(2)$ symmetry, with the conserved $S^z$
taking the role of boson ``charge''.  This situation is not uncommon,
as it is realized whenever an approximately isotropic magnet is
subjected to a uniform magnetic (Zeeman) field. We will focus on this realization here.

\begin{figure}%[!hbp]
\begin{center}
\ifig[height=2.0in]{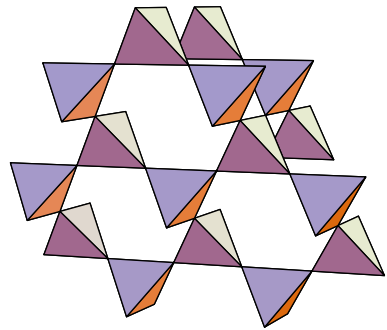}
\caption{(Color online.) The pyrochlore lattice structures, shown as a network of corner sharing tetrahedra. 
The atoms occupy the corners of the teterahedra.}
\label{fig:pyrochlore}
\end{center}
\end{figure}

In this paper, we will apply an analogous set of ideas to bosons/spins
on the three-dimensional pyrochlore lattice.  The pyrochlore lattice,
consisting of corner-sharing tetrahedra (Fig.~\ref{fig:pyrochlore}),
takes a central role in the study of geometrical frustration in three
dimensions.  A number of materials, in which electronic and/or spin
degrees of freedom reside on this lattice, have been intensely studied
in this light in recent years.  Theoretically, the Heisenberg
antiferromagnet on the pyrochlore lattice is interesting as a candidate
``spin liquid'' (Ref.~\onlinecite{Chalker:prl98,Hermele:prb04}), in which the fluctuations -- thermal or quantum --
amongst frustration-induced degenerate quasi-ground states prevent the
occurrence of long-range magnetic order at temperatures well below the
Curie-Weiss temperature, possibly all the way down to $T=0$.  Quantum
spin liquid states can sustain anomalous spin-$1/2$ {\sl spinon}
excitations~\cite{Anderson:87}, forbidden in conventional phases of matter.  

We therefore choose to take as our proximate liquid phase not a
superfluid (or magnetically ordered phase), but instead a particular
quantum spin liquid, a so-called ``U(1)'' spin liquid state (see Ref.~\onlinecite{Wen:prb02} and 
references therein).  A general feature (e.g. on different varieties of lattices) of such states is that
they exhibit an {\sl emergent electromagnetism}, i.e. they support an
``artificial photon'', and excitations can carry {\sl emergent} U(1)
electric and magnetic gauge charges (see Ref.~\onlinecite{Hermele:prb04,Moessner:prb03,Foerster:plb80} 
and references therein).  The 
spinons carry the elementary
quanta of the electric charge (of both positive and negative sign in two
species of spinon).  Another, gapped, topological excitation, a {\sl
  monopole}, carries the dual magnetic gauge charge.  A transition out
of the spin liquid state to a state without broken continuous symmetries
is generally described as a condensation of these monopoles.  The nature
of such a transition, and of the proximate spatially-ordered states
occuring on the other side of the transition is determined by the
monopole PSG.  The ingredients determining this PSG, as explored on the
cubic lattice in Refs.\onlinecite{Motrunich:prb05,YBKim:04}, are the
lattice symmetries and the values of some conserved ``background'' U(1)
gauge charges, which characterize different U(1) liquid states.  

To fix these background charges, we will focus on a specific model
containing a U(1) spin liquid phase on the pyrochlore lattice.  In
Ref.\onlinecite{Bergman:prl05}, this model was argued to describe the
physics on the magnetization plateaus observed recently in
\cdaf\,, \hgaf\,\cite{Ueda:prl05,Ueda_unpub}, which are spin-$3/2$
antiferromagnets with this lattice structure.  In particular, the model
presumes a local constraint (which may be understood as the restriction
to the classical ground state subspace) of three Cr spins fully
polarized ($S^z=+3/2$) along the applied field, and another fully
polarized antiparallel to it ($S^z=-3/2$), {\sl on each
  tetrahedron}\cite{Penc:prl04,Bergman:prl05}. We refer to this
condition as the ``3:1'' constraint.  The model of
Ref.\onlinecite{Bergman:prl05} arises as an effective Hamiltonian in
this constrained subspace, and takes the approximate form:
\begin{eqnarray}
  \label{eq:Happrox}
  {\mathcal H}_{\textrm{QDM}} & = & V \sum_P \left(
  \ket{\hexagon_A} \bra{\hexagon_A} + \ket{\hexagon_B} \bra{\hexagon_B}
\right) \nonumber \\ &&  - K \sum_P \left( \ket{\hexagon_A} \bra{\hexagon_B} + h.c.
\right), 
\end{eqnarray}
where $\sum_P$ indicates a sum over all hexagonal plaquettes on the
pyrochlore lattice, and $\ket{\hexagon_A},\ket{\hexagon_B}$ are specific
states with alternating majority and minority spins
($|\!\!\uparrow\downarrow\uparrow\downarrow\uparrow\downarrow\rangle$ and
$|\!\!\downarrow\uparrow\downarrow\uparrow\downarrow\uparrow\rangle$) on the
given plaquette.  This model is exactly equivalent to a number of other
models in the theoretical literature.  First, it can be mapped directly
to a quantum dimer model on the diamond lattice, the diamond lattice
sites being centers of pyrochlore tetrahedra (see Sec.~\ref{sec:model}).  A number of
such dimer models have been considered in the literature~\cite{Misguich:prl02,Misguich:prb05,Ralko:05,Moessner:prb01,Moessner:prl01,Huse}.  Second, the
dimer model in turn can be rewritten as a particular compact U(1) gauge
theory.  The 3:1 constraint of the spin model maps directly to the
background charge of this gauge theory.  In this way, the essential
ingredients fixing the monopole PSG are determined.  A systematic
analysis of the spatially-ordered states proximate to the spin liquid is
therefore possible, and is the main subject pursued in this paper.

More microscopically, it is possible to show that the spin/dimer model
of Eq.~(\ref{eq:Happrox}) indeed exhibits a U(1) spin liquid ground
state when the dimensionless parameter $v=V/K$ satisfies $v_c < v < 1$.
This argument, analogous to the ones in Refs.\onlinecite{Hermele:prb04,Huse},
is described in Sec.~\ref{sec:phase-diagr-quant}.
The critical coupling $v_c$ is not known, but based on numerical
analysis of other similar models probably satisfies $v_c> -0.5$ or so
\cite{Moessner:prl01,Moessner:prb01,Huse,RK:prl88}.
%\cite{QDM,RK:prl88}
For the application to \cdaf\,, \hgaf\,, it was estimated in
Ref.\onlinecite{Bergman:prl05} that $v \approx -1.2 < v_c$.  The nature
of the ground state in that case may perhaps be more accurately
understood by extrapolation from the limit $v\rightarrow -\infty$.  The
ground state can be determined classically in that limit, and in
Ref.\onlinecite{Bergman:prl05} was found to exhibit a particular spatial
ordering pattern with a quadrupled unit cell.  It can be understood by
order-by-disorder\cite{Henley:prl89,Henley:cjp01,Villain} reasoning as the classical state
with the most possible ``resonances'' -- off-diagonal quantum moves via
the $K$ term in Eq.~(\ref{eq:Happrox}) -- to other state.  We therefore
refer to it as the ``R'' state.  

The alternate approach, which we pursue here, is to approach the physical
limit from the spin liquid state, asking which ordering pattern emerges
from the PSG analysis.  Remarkably, we find that {\sl the simplest
  possible ordered state proximate to the U(1) spin liquid is the R
  state}!  This suggests the possibility that the R state in the
physical limit may be close to a phase transition to the spin
liquid state.  The analysis of this paper provides an analytical
framework for such a transition, both of quantum and of thermal nature.

\begin{figure}[htb]
\begin{center}
%\vskip+6mm
%\centerline{\ifig[width=1.3in, height=1.0in]{fig2}}
\ifig[width=1.7in, height=1.7in]{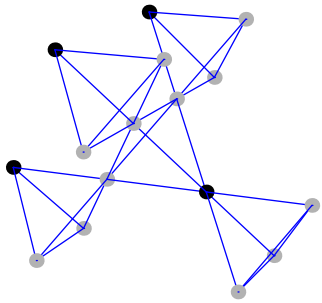}
\caption{ The simplest ordered state consistent with the
  3:1 proportion of majority to minority sites, which does {\sl
    not} exhibit an enlarged unit cell, and has only a four-fold
  degeneracy.  In the spin language, this state has three spins
  aligned with the field and one anti-aligned, so we will denote this
  as a ``ferrimagnetic'' state, analogous to the ferrimagnetic state
  found in two-dimensional triangular lattice antiferromagnets in a
  field.  This state is {\sl not} generically proximate to the $U(1)$ spin liquid. }
\label{fig:ion_valence}
\end{center}
\end{figure}

Apart from the simplest R state, a number of ordered phases come out
of this analysis, and are shown in
Figs.~\ref{fig:dense1}-\ref{fig:dense4D}.  A salient feature of all
these phases is that they exhibit an enlarged unit cell relative to
that of the original pyrochlore lattice.  Interestingly, this set of
phases therefore {\sl does not} contain the simplest ``ferrimagnetic''
ordered state illustrated in Fig.~\ref{fig:ion_valence}, in which the
unit cell is {\sl not} enlarged though the point group symmetry of the
crystal is lowered.  This indicates that the ground states selected
out of the classically degenerate ground state manifold are
identifiably different from those preferred by other degeneracy
breaking mechanisms, such as e.g antiferromagnetic {\sl
  second-neighbor} exchange coupling.  Identification of the precise
nature of the ordered state in experiments therefore indirectly gives
useful information on the importance of quantum fluctuations.

The remainder of the paper is organized in the following way.  In
Sec.~\ref{sec:model} we describe our theoretical model (lattice QED).
In Sec.~\ref{sec:duality} we illustrate and motivate the dual
transformation we make on the Hamiltonian introduced in the previous
section.  In Sec.~\ref{sec:action} we derive and present the effective
action for the monopole degrees of freedom that appear in the dual
theory. We then study the action in the mean-field approximation in
Sec.~\ref{sec:mean-field} and present the resulting charge ordered
phases in Sec.~\ref{sec:charge-order-patt}.  In Sec.~\ref{sec:rg-analysis}
we carry out a renormalization group analysis of the action.	Finally in
Sec.~\ref{sec:conclusions}, we conclude with a discussion of the phase
diagram and critical phenomena in the spin/dimer model at non-zero
temperatures.  Important but lengthy formulas and results are given in
the appendices.

\section{Theoretical Model: From Spins to Compact Lattice QED}
\label{sec:model}

In this section, we reformulate the 3:1 quantum
spin/dimer model of Eq.~(\ref{eq:Happrox}) as a lattice U(1) gauge
theory, and describe its phase diagram, which contains both spin
liquid and ordered states.  This Hamiltonian has a
single dimensionless parameter, $v=V/K$, so the zero temperature phase
diagram is entirely determined by $v$ (we fix $K>0$ by convention --
its sign can be changed by a suitable canonical transformation, and
has no significance).  We discuss the structure of this phase
diagram.  

\subsection{Equivalence to compact lattice QED}
\label{sec:equiv-comp-latt}

\begin{figure}[!hbp]
\begin{center}
%\vskip+6mm
\ifig[height=2.0in]{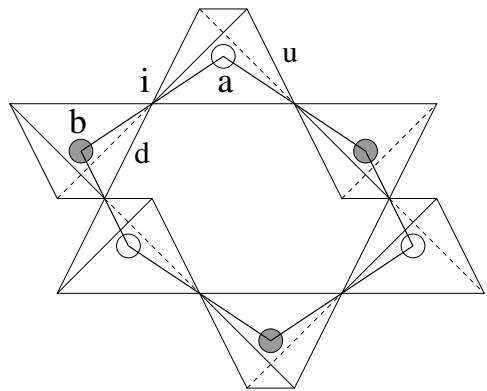}
\caption{Section of a pyrochlore lattice. The pyrochlore sites are denoted by $i$ and the tetrahedra
are identified by $a,b$ labels. 
Drawing links between the tetrahedron centers forms the links of a diamond lattice, with the tetrahedron centers
corresponding to the sites of the diamond lattice.
The figure indicates the bipartite nature of the diamond lattice, which is evident in the notion of up/down pointing tetrahedra as indicated by the 
labels u/d.}
%\vskip-4mm
\label{fig:pyro_diamond}
\end{center}
\end{figure}

First, it is useful to discuss how the model can be cast into a
lattice U(1) gauge theory.  A reader not familiar with lattice QED may
wish to consult the review by Kogut, Ref.~\onlinecite{Kogut:rmp79}.
As mentioned in the introduction, it is first convenient to pass from
the pyrochlore to the diamond lattice.  This is accomplished by
focusing on the centers of the tetrahedra (labeled by the sites $a$ and
$b$ in Fig.~\ref{fig:pyro_diamond}) that make up the pyrochlore.  The
centers of these tetrahedra make up a diamond lattice.  Each site on the
pyrochlore lattice connects two nearest neighbor tetrahedra, and can
be identified with a link between the centers of the two tetrahedra.
Thinking in terms of the centers of the tetrahedra as the sites of a
new lattice, the spins sit on the links of a diamond lattice.  The
spin states may therefore be regarded as dimer coverings of the links
of the diamond lattice, and the effective Hamiltonian as a quantum dimer model.

The gauge nature of the problem is simply a consequence of the local
3:1 constraint on each pyrochlore tetrahedron, or equivalently, that
each diamond site is covered by a single dimer.  To map this onto a
more conventional gauge theory, we note further that the diamond
lattice is bipartite, so we can define ``up'' ($\u$) and ``down''
($\d$) sub-lattices corresponding to neighboring tetrahedra in the
original pyrochlore lattice, as illustrated in
Fig.~\ref{fig:pyro_diamond}.  We can define thereby a discrete {\sl
  oriented} electric field variable on the diamond lattice links,
equal to zero if the corresponding pyrochlore site is a majority site,
and equal to $\pm 1$ if the pyrochlore site is a minority one,
choosing the field to always point from the up to the down diamond
sublattice.  We specify the spin/dimer configurations on the
pyrochlore by a discrete variable $\hat{n}_i$ ($i$ is a pyrochlore
lattice site), such that
\begin{equation}
  \label{eq:ndef}
  \hat{n}_i = \left\{\begin{array}{cc} 0 & \textrm{majority site,} \\ 1
      & \textrm{minority site.} \end{array} \right.
\end{equation}

Mathematically, for the site $i$ lying on the diamond link $ab$, 
\begin{align}
\label{eq:E_ab}
E_{a b} = \epsilon_a \hat{n}_{a b}\;, 
\end{align}
where 
\be
\label{eq:e_a}
\epsilon_a = \left\{
\begin{array}{ll}
+1 & , \, a \in \u 
\\
-1 & , \, a \in \d
\end{array}
\right.\;.
\ee
The electric field direction can be
identified by the index ordering that gives it a positive value, as in  
Fig.~\ref{fig:pyro_diamond}.

The model we construct requires that each tetrahedron must include $1$
minority site, 
\be\label{1boson}
\hat N_a \equiv \sum_{i \in a} {\hat n}_i = 1\;,
\ee
where the label $a$ identifies the various tetrahedra, and 
$i $ is summed over pyrochlore lattice sites on tetrahedron $a$.
See Fig.~\ref{fig:pyro_diamond} for an illustration.
This 3:1 constraint on the pyrochlore lattice in the QED formulation
maps directly into 
\be\label{Gauss} \epsilon_a = \divo \vec E\;,
\ee which is rather reminiscent of Gauss' law, where we have used
the lattice divergence $ \divo \vec E = \sum_b E_{a b} $. The ``charge
distribution'' in this picture is that of alternating positive and
negative charges on the diamond lattice. Positive charges sit on the
$\u$ sub-lattice, and negative charges sit on the $\d$ sub-lattice. 

A further consequence of this mapping is the presence of {\sl global}
topological charges, which are conserved in periodic (or infinite)
systems.  In particular, if one draws any surface not passing directly
through diamond sites, the net electric flux through this surface
is conserved.  If this surface is compact and closed, this gives no
additional information beyond the Gauss' law of Eq.(\ref{Gauss}).
However, one may also consider non-compact surfaces that extend across
the entire sample, and the electric flux through such surfaces is not
determined by Gauss' law.  A sufficient set of surfaces are the four
non-parallel flat planes containing two-dimensional triangular
lattices of pyrochlore sites (any two such parallel planes have the
same electric flux).  We denote the corresponding four electric fluxes
by a four-vector $(E_1,E_2,E_3,E_4)$.  The fluxes may be chosen
positive, and can in principle take any integer value from $0 \leq E_i
\leq N_\triangle$, where $N_\triangle$ is the number of triangular
sites in the plane.  By the 3:1 (Gauss' law) constraint, the sum
$E_1+E_2+E_3+E_4=N_\triangle$ is fixed.  The electric flux sector
containing the ground state varies with $v$.

It is conceptually useful to also map the full Hamiltonian of the
model to a form more familiar in lattice QED.  To do so, we must
introduce the phase operator $\hat{\phi}_i$ conjugate to the number
operator $\hat{n}_i$, satisfying
\begin{equation}
\Big[
{\hat \phi}_{j}  , \, \hat n_{i}
\Big] = +i \, \delta_{ji}\;,
\end{equation}
where $\delta_{ij}$ is the Kronecker delta function.  The operator
$e^{+i \phi_j}$ creates a minority site at site $j$, and using it we
can construct any hopping term we wish for minority sites (down
spins).  With the canonical ``rotor'' variables
$\hat{n}_i$, and $\hat{\phi}_i$, in principle an infinite set of number
states with all integer eigenvalues of $\hat{n}_i$ are allowed.  To
faithfully represent the original spin/dimer model, therefore, we will
include a large term $U$ in the Hamiltonian which, in the limit
$U\rightarrow \infty$, restricts the site occupancies to $\hat{n}_i=0,1$ as desired.
We thereby obtain \be
\label{bosonH}
\begin{split}
{\mathcal H} =  &\frac{U}{2} \sum_{i} \hat n_i \left(\hat  n_i - 1  \right) 
+ U_t \sum_a ({\hat N}_a - 1)^2
\\ + & V \sum_{\substack{\hexagon }}
\Big[\delta_{n_1,1}\delta_{n_2,0}\delta_{n_3,1}\delta_{n_4,0}\delta_{n_5,1}\delta_{n_6,0}+
(n_i \leftrightarrow 1-n_i)\Big]
\\ - & 
\frac{K}{2} \sum_{\substack{\hexagon }} \Big[
e^{+i\left(
\phi_{1}-\phi_{2}+\phi_{3}-\phi_{4}+\phi_{5}-\phi_{6}
\right)} + h.c \Big]
\;.
\end{split}
\ee
Here $\hexagon$ denotes the hexagonal plaquettes on the pyrochlore
lattice, and the indices on $\phi$ enumerate the site (links) on the
hexagon. The constraint operators $\hat{N}_a$ commute with ${\cal H}$ by
construction, so for sufficiently large $U_t$ the ground state will
indeed satisfy Eq~(\ref{1boson}).  Moreover, when the constraint is
enforced, the $U_t$ term plays no further role.  Formally, we are
principally interested in the limit $U/K \rightarrow \infty$, as
described above.

\begin{figure}[!hbp]
\begin{center}
%\vskip+6mm
\ifig[height=1.3in]{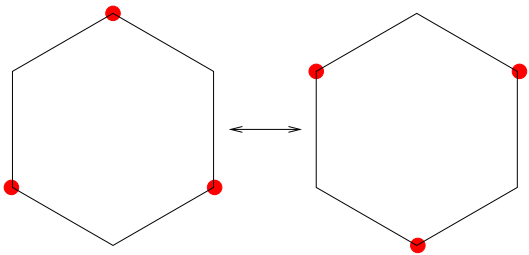}
        \caption{(Color online.) Ring exchange hopping on a hexagonal plaquette. 
        The hopping flips between A and B type plaquettes, and
        is the only simple hopping event
        that preserve the 1 boson per tetrahedron constraint.}
        %\vskip-4mm
\label{fig:hopping}

%        \centering
%                \ifig[width=2.0in]{MayFig_ringAB.eps}
%        \caption{(Color online.) Ring exchange hopping flips between A and B type plaquettes.}
%        \label{fig:MayFig_ringAB}
\end{center}
\end{figure}

The $K$ term when rewritten in this way appears as a rather
complicated-looking multi-particle hopping amplitude.  In fact, this
form is actually the simplest one allowed by the constraint
\eqref{1boson}. The hopping of a down spin from one lattice site to
another can be decomposed into a series of hops along nearest neighbor
links, so it is sufficient to analyze the simplest allowed moves. In
general, the hopping on a nearest-neighbor link will
violate \eqref{1boson} on 2 separate tetrahedra. Thus, any series of
such hopping events will do the same. Analysis shows that it is only
possible to hop between tetrahedra along \emph{closed}
contours. On the pyrochlore lattice the smallest closed contours are
hexagonal plaquettes. Any other closed contour on the lattice can be
constructed from these minimal moves, so we shall consider exclusively
this ``ring exchange hopping'' on the hexagonal plaquettes. Such moves
are illustrated in Fig.~\ref{fig:hopping}.

%The Hamiltonian can also be expressed more compactly as
%\be\label{bosonH_flip}
%\begin{split}
%{\mathcal H} & =  \frac{U}{2} \sum_{i} \hat n_i \left( \hat n_i - 1  \right) 
%+ V \sum_a (N_a - 1)^2
%\\ &
%-
%\frac{K}{2} \sum_{\substack{\hexagon }} \Big[
%\ket{\hexagon_A} \bra{\hexagon_B}
% + h.c \Big]\;,
%\end{split}
%\ee 
%where $\hexagon_{A,B}$ denote the 2 different possible states of a
%ring with 3 bosons on it, as illustrated in Fig.~\ref{fig:hopping}.

This Hamiltonian can now be re-expressed as a lattice gauge theory.
Analogously to the definition of the electric field, we define the
vector potential as
\begin{equation}\label{eq:E_A}
\A_{a b} = \epsilon_a \phi_{a b}\;.
\end{equation}
The electromagnetic variables introduced above obey the same canonical
commutation relations as $\hat \phi$ and $\hat n$, since
$\epsilon_a^2=1$, 
\be\label{eq:A_ab}
\Big[
\A_{a b} , \, E_{a b}
\Big] = i\;,
\ee
on the same link, and the commutator is $0$ on different links. Note
that from their definitions, $E_{ab}$ is integer-valued (and in
particular $=0,\pm 1$) and $A_{ab}$ is a $2\pi$-periodic phase variable
since the operator $e^{i\hat \phi}$ creates a particle.

In the new variables the ring exchange term becomes
\be
\bs &
e^{+i\left(
\A_{1} + \A_{2} + \A_{3} + \A_{4} + \A_{5} + \A_{6}
\right) } + h.c. = \\ &
e^{+i \oint {\vec \A} \cdot d{\vec \ell} } + h.c. = 
2 \, \cos{(
\curl{\vec \A}
)_{\hexagon}}\;,
\end{split}
\ee
where we have introduced the lattice curl:
\be\label{eq:curl-defn}
(\curl {\vec \A})_{\hexagon} = 
\sum_{{\vec r}{\vec r}' \in \hexagon}\!\!\!\!\!\!\!\!\!\!\!\circlearrowleft
\; {\vec \A}_{{\vec r}{\vec r'}}\;.
\ee
In this form, the previously complicated-looking form of the $K$ term
becomes transparent.

After substituting the new variables, the Hamiltonian \eqref{bosonH} 
takes the form
\begin{eqnarray}\label{QED1}
{\mathcal H} & = & const. + \frac{U}{2} \sum_{\substack{\langle a,b\rangle }} 
\left( E_{a b} - \frac{\epsilon_a}{2}  \right)^{2} - 
K \sum_{\substack{\hexagon }} \cos \left( \curl {\vec \A} \right) \nonumber
\\
& & + V  \sum_{\substack{\hexagon }} \left(\delta_{\curl {\vec E}, 3} +
\delta_{\curl {\vec E}, -3} \right),
\end{eqnarray}
where the constant is a result of reorganizing the first term of 
\eqref{bosonH}  into a quadratic form. By identifying the curl of the
vector potential with a 
magnetic field running through the plaquettes, 
${\vec B} = \curl {\vec \A}$,
\begin{eqnarray}\label{QED2}
{\mathcal H} &  = &  const. + \frac{U}{2} \sum_{\substack{\langle
    a,b\rangle }}  \left( E_{a b} - \frac{\epsilon_a}{2}  \right)^{2} - 
K \sum_{\substack{\hexagon }} \cos{\vec B} \nonumber \\
& & + V  \sum_{\substack{\hexagon }} \left(\delta_{\curl {\vec E}, 3} +
\delta_{\curl {\vec E}, -3} \right).
\end{eqnarray}
After these manipulations, the spin/dimer Hamiltonian has been
formulated as a {\it{compact}} Quantum Electrodynamics (QED) in 3+1
dimensions.  Eq.~(\ref{QED2}) is very similar to the standard form of
compact QED, but does differ from it by the presence of the rather
ugly $V$ term and the modification of the $E^2$ term by the
$\epsilon_a/2$ ``background field''.  Please recall also that we are
expected to take the limit $U\rightarrow \infty$ to recover the
spin/dimer model.  Despite the differences, Eq.~(\ref{QED2}) does
share all the same internal symmetries as the more conventional QED
form.  It is therefore expected to share the same properties in
regimes where universality is mandated.

\subsection{Phase diagram of the quantum spin/dimer model}
\label{sec:phase-diagr-quant}

Let us now return to the microscopic form of the spin/dimer model, and
the question of the phase diagram.  We will employ the QED formulation
where useful in this analysis.  For $v = \frac{V}{K} \rightarrow -\infty$, the
off-diagonal $K$ term can be neglected, and ground state is determined
by minimizing the (negative) $V$ term over {\sl classical} spin/dimer
configurations.  The solution is the R state, shown in Fig.~\ref{fig:dense1}.  This state has only a
discrete degeneracy, and is separated from other excited states by a gap
(of $O(V)$), so it is expected to be stable to perturbation theory in
$K$.  Therefore the R state is the ground state for $v<v_{c1}$, with
some $v_{c1}<1$.  In the R state, the electric flux is equally
divided, $E_i/N_\triangle = 1/4$ apart from $O(1/N_\triangle)$
corrections for some frustrated boundary conditions.

For $v >1$, the ground state
can be found by rewriting the Hamiltonian as follows:
\begin{eqnarray}\label{eq:projectors}
    {\mathcal H}_{\textrm{QDM}} & = & K \sum_P (
  \ket{\hexagon_A}-\ket{\hexagon_B})(\bra{\hexagon_A} -
  \bra{\hexagon_B})  \nonumber \\ &  + & (v-1)K \sum_P \left( \ket{\hexagon_A} \bra{\hexagon_A} + \ket{\hexagon_B} \bra{\hexagon_B}\right).
\end{eqnarray}
In Eq.(\ref{eq:projectors}) $\mathcal{H}_{\textrm{QDM}}$ has been
expressed as a sum of positive semi--definite projection operators, with
coefficients that are all positive for $v>1$.  Therefore the energy is
bounded below by zero, and any zero energy state is a ground state.
In particular, any classical state that contains {\sl no} A or B
hexagons is automatically a ground state.  As a simple example,
consider the ``ferrimagnetic'' state, with no enlargement of the unit
cell but broken rotational symmetry.  It is described as follows.  On
the diamond lattice, each vertex has 4 links emanating from it, which
we label by $\mu = 0,1,2,3$. The ferrimagnetic state (oriented along a
specific $\mu$-direction in space) has the same 3:1 arrangement on all
identical tetrahedra.  Explicitly, it can be written \be
\ket{\textrm{Ferri}}_{\mu} = \prod_{a \in \u} \ket{n_{a,\mu} =1}\;,
\ee where the product is taken over the $\u$ sublattice of tetrahedra
shown in Fig.~\ref{fig:pyro_diamond}.  This state clearly obeys the
3:1 constraint \eqref{1boson} -- one minority site on the $\mu$ corner of each
tetrahedron.  And moreover, it contains no A or B plaquettes, and
hence is a zero energy ground state for $v>1$.  The ferrimagnetic
state is in the ``furthest'' topological sector from the R state, with
$E_i = (N_\triangle,0,0,0)$ (and permutations).

The ground states in this regime
are, however, highly degenerate, and many other classical
configurations are possible.  All such states are ``frozen'' in that, for
any value of $v$, they are exact eigenstates with trivial dynamics,
being annihilated by the off-diagonal $K$ term in
Eq.~(\ref{eq:Happrox}).

At the point $v=1$, Eq.~(\ref{eq:projectors}) simplifies (the last
term dropping out), but remains the sum of positive semidefinite
projectors.  This is the so-called Rokhsar-Kivelson (RK) point.  The
ground state space is enlarged, and contains now states in
addition to the frozen ones.  In particular, many configurations
containing A and B hexagons are now allowed, provided the first
projector in Eq.~(\ref{eq:projectors}) annihilates the quantum state.
One construction of this type is especially simple.  Take a {\sl
  uniform} superposition of {\sl all} possible spin configurations,
ignoring the 3:1 constraint.  Now project this onto the 3:1 manifold.
This is known as the RK wavefunction. It can be further broken
into substates, e.g. by projecting out all the frozen states.  It can
also be projected into any of the electric flux sectors.

On reducing $v$ to values slightly {\sl below} unity, the second term
in Eq.~(\ref{eq:projectors}) becomes again non-zero but negative
semi-definite.  This indicates configurations with A and B hexagons
are now {\sl preferred} in the ground state.  The frozen states are
then highly excited and energetically unfavorable.  Instead, by an
application of the reasoning of Ref.\onlinecite{Hermele:prb04}, the
ground state can be argued to be a U(1) spin liquid state.

We summarize this argument, and the nature of the U(1) state.  It is
by now well-known that RK wavefunctions for dimer models on bipartite
lattices (like the diamond) display power-law equal-time correlations.
These correlations may be understood as arising from the 3:1
constraint.  In particular, the calculation of equal-time correlations
in the RK state reduces to a problem of three dimensional classical
statistical mechanics: performing a statistical average over discrete
electric field configurations subject to the Gauss' law constraint of
Eq.~(\ref{Gauss}).  It turns out that the long-distance behavior of
these correlations is captured simply by taking an effective classical
free energy density proportional to $|\vec E|^2$ and treating $\vec E$
as a continuous Gaussian (but constrained) variable.  The resulting
correlations have a ``dipolar'' power-law
form\cite{Hermele:prb04,Isakov}.

By an argument originally due to Henley\cite{Henley:cjp01}, these
power laws can be understood in the quantum theory as follows.
Evidently, the discreteness of the electric field is unimportant at
the RK point.  Consequently, it is natural to treat $\vec{E}$ and
$\vec{B}$ as continuous, write an effective action quadratic in these
fields, and apply the Gauss' law constraint.  However, there is an
additional feature dictated by another peculiarity of the model.  As
we have pointed out already, all possible electric flux sectors are
degenerate at the RK point.  This implies that there should be no cost
in the energy to shift $\vec{E}$ by a uniform constant (at the RK
point).  Henley's argument therefore indicates an appropriate
effective Hamiltonian density is
\begin{eqnarray}
  \label{eq:lagrange}
  \mathcal{H}_{\rm eff} & = & a |\curl \vec{E}|^2 + b |\vec{B}|^2 +  \alpha
  (1-v) |\vec{E}|^2 .
\end{eqnarray}
The last $|E|^2$ term must vanish at the RK point, but is expected to
become non-zero if one perturbs away from it.  For $v=1$, this form
can be shown to precisely reproduce the microscopically calculated
correlations of the RK wavefunction, with specific constants $a,b$.
For $v$ larger than one, the negative $|\vec{E}|^2$ term favors the
sectors with ``large electric flux'', i.e. the frozen states, as
expected, and $\vec{E}$ itself develops a non-zero expectation value.
For $v$ slightly less than one, the positive $|\vec{E}|^2$ term
instead favors the ``minimal electric flux'' sector with $E_i =
N_\triangle/4$.  

At low energies, for $v\lesssim 1$, therefore, it is expected that the
$a$ term above can be dropped in favor of the $\alpha$ term, making
the effective Hamiltonian simply that of the usual {\sl non-compact}
QED.  This indicates the system is in the ``Coulomb phase'' of the
gauge theory, which has the usual properties expected of 3+1
dimensional electrodynamics.  In particular, unit {\sl test} gauge charges
can be introduced and interact via bounded $1/r$ Coulomb potentials.
Such a gauge charge corresponds in the original pyrochlore magnet to a
``spinon'' excitation with fractional spin $\pm 3/2$.  Thus the
Coulomb phase is indeed a spin liquid, and in respect of the U(1)
gauge structure, this state is called a U(1) spin liquid.  

A key difference from standard ``non--compact'' QED {\sl does} appear at non-zero
energies, as a consequence of the fact that the ${\vec B}$ is defined
modulo $2\pi$. The divergence of this magnetic field is the sum of
magnetic field values coming out of the plaquettes enclosing one cell
in the diamond lattice.  With the modulo $2\pi$ redundancy,
configurations with $\divo{\vec B}$ an integer multiple of $2\pi$ are
allowed.  Thus, compact QED allows magnetic {\it{monopoles}} with
quantized magnetic charge.  These have a finite energy cost and are
gapped excitations in the U(1) spin liquid.

Although the Coulomb phase emerges in a non-trivial way from the
spin/dimer model in the vicinity of the RK point, we can mimic its low
energy physics more simply.  In particular, the same phase is obtained
from Eq.~(\ref{QED2}) by {\sl dropping} the diagonal $V$ term, and
taking non-zero but finite $U$ (instead of $U\rightarrow \infty$).
This gives a ``softened'' model with the same universal properties as
the original spin/dimer model.  It is a remarkable fact that, by the
preceding arguments, these two ``sins'' compensate themselves and give
the proper behavior of the original spin/dimer model for $v \lesssim
1$, in the Coulomb phase.

\section{Duality and monopole formulation}
\label{sec:duality}

On reducing $v$ from values just below one, eventually the spin/dimer
model must undergo a transition out of the spin liquid state.  The
resulting state can be mimicked in the softened model by increasing
$U/K$.  The eventual outcome can be understood as follows.  For $U \gg K$ the
``magnetic field'' term is subdominant, and so the electric field is a
good quantum number.  This limit is somewhat complicated in
Eq.~(\ref{QED2}), because the $U$ term selects {\sl two} degenerate
values $E_{ab}=0,\epsilon_a$ as low energy states of each bond.
Indeed this recovers the original effective spin-$3/2$
model, which is of course still non-trivial.  However, one can readily
understand in this limit the basic nature of the other phases of the
theory.  To do so, we imagine generalizing the $U$ term to include
electric-field interactions on nearby bonds.  This will generally
break the large $U$ degeneracy in favor of some particular global
arrangement of $E_{ab}=0,\epsilon_a$ values.  Because of the
discreteness of $E_{ab}$, deviations from this ground state are
likewise discrete, any local rearrangement of the pattern results in a
non-zero increase in energy, i.e. there is an energy gap.
Gauge-neutral excitations are created in this way.  Either a set of
$E_{ab}$ fields are modified along links forming a closed curve, or a
pair of diamond lattice sites (on which the Gauss' law constraint is
violated) is created, with a modified path of $E_{ab}$ fields
connecting them.  The latter corresponds to a particle-antiparticle
pair, and costs an energy proportional to the length of the path.
Hence the pair itself is bound, the bound state being gauge neutral.
The individual gauge-charged excitations are said to be {\sl confined}
by the linear potential between them.  The confined phase corresponds
to having the $\hat{n}_i$ operators certain in \eqref{bosonH} -- a
Mott insulating phase, with some sort of diagonal ordering.  In contrast
to the Coulomb phase, the $\vec{B}$ field in the confining phase is
very strongly fluctuating, so monopoles are no longer good
excitations.  In fact, it is appropriate to think of the confined
phase as a Coulomb phase which has been destroyed by {\sl Bose
  condensation} of monopoles.  The presence of a delocalized monopole
condensate can be thought of as leading to strong (gauge) magnetic
fluctuations in the ground state.

We therefore wish to reformulate the lattice QED model so that the
monopole excitations of the Coulomb phase are explicit.  Here we
follow Hermele {\sl et al.}\cite{Hermele:prb04} with slight differences.  As
the reader will recall, the electric and magnetic fields in Maxwell's
equations are dual when there are no {\it{charges}} or {\it{currents}}
present.  While there are no currents in our system, Eq.~\eqref{Gauss}
shows we do have a charge distribution. The duality transformation is
\begin{align}
{\vec E}_{ab} & = \curl {\vec \alpha} + {\vec e}_{ab}^{(0)} \;,\\
{\vec B} & = \curl {\vec \A}\;.
\end{align}
We have thus introduced an explicit operator for the magnetic field
${\vec B}$ and an ``electric vector potential'' ${\vec \alpha}$, whose
exponential creates a `magnetic field' since an exponential of $\vec \A$
creates an ``electric field'', via \eqref{eq:E_A},\eqref{eq:e_a},\eqref{eq:E_ab},\eqref{eq:A_ab}.  Here
${\vec e}^{(0)}$ is a classical electric field created by the charge
distribution \eqref{Gauss}, and 
\be \divo {\vec e}^{(0)} = \divo {\vec
  E} = \epsilon_a.  
\ee 
It is convenient to choose ${\vec
  e}_{ab}^{(0)}$ to be integer-valued, so that curl${\vec\alpha}$ may also
be taken integer-valued.  A simple choice is to take the classical
configuration corresponding to one of the 3:1 states, e.g. just
$e^{(0)}_{a,a+\mu} = \epsilon_a \delta_{\mu 0}$.

\begin{figure}[!hbp]
\begin{center}
%\vskip+6mm
\ifig[height=3.0in]{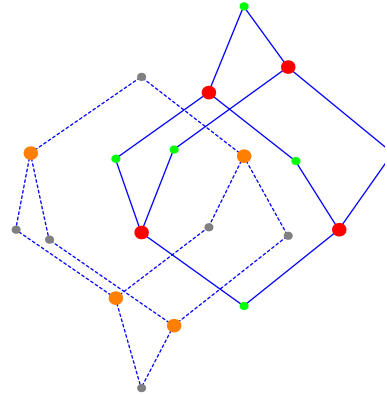}
\caption{(Color online.) Direct and dual diamond lattices are dual to one another. The plaquettes of one 
lattice correspond to the links of its dual lattice.}
%\vskip-4mm
\label{fig:dual_lattice}
\end{center}
\end{figure}

The Hamiltonian in the dual language takes the form
\be\label{dual_QED}
{\mathcal H} = \frac{U}{2} \sum_{\substack{\hexagon}}
\left( \curl {\vec \alpha} + e^{(0)}  - \frac{\epsilon_a}{2}  \right)^{2} -
K \sum_{\substack{r,r'}} \cos{\vec B}\;,
\ee
where we denote the sites of the dual lattice by $r$, and its links by $r,r'$. The hexagons now denote the
plaquettes of the dual lattice. (Note the change in the summation subscripts on both the first and second terms.)

The dual fields obey the canonical commutation relations
\be\label{dual_commutator}
\Big[
B_{r,r'} , \, \alpha_{r,r'} 
\Big] = +i\;,
\ee
and the commutator vanishes for different links. The new fields are
once again conjugate variables.  $B_{r,r'}$ is defined modulo $2\pi$ -
an angular variable, and the $\alpha_{r,r'}$ variable is integer
valued. 

Standard manipulations can now be used to ``soften'' the inconvenient
integer constraint on $\alpha_{r,r'}$, remove the periodicity of
$B_{r,r'}$, and make the monopole variables explicit.  The reader is
referred to Refs.\onlinecite{Kyoto:04,Motrunich:prb05}\ and references
therein for details.  These manipulations are inexact, but do not
change the structure of the phase diagram in the vicinity of the
transition from the Coulomb to confining phase.  One obtains 
\begin{equation}
  \label{gaugedH}
  \begin{split}
    {\mathcal H} = \frac{U}{2} \sum_{\substack{\hexagon}} & \left( \curl
      {\vec \alpha} -\overline{e}\right)^{2} +
    \frac{K}{2} \sum_{\substack{r,r'}} {\vec B}^{2} \\ - & w \sum_{r,r'}
    \cos [\theta_r- \theta_{r'} -2\pi  \alpha_{r,r'}] \;,
  \end{split}
\end{equation}
where now $\alpha_{r,r'}$ and ${\vec B}$ are real variables.  In
Eq.~(\ref{gaugedH}), one may freely shift $\overline{e}$ by a
gradient, changing only the overall zero of energy, since such a
gradient does not couple to $\curl \vec\alpha$.  We have used this
freedom to modify the original $e^{(0)}+\epsilon_a/2$ terms to
\begin{equation}
  \label{eq:ebar}
  \overline{e}_{a,a+\mu} = \epsilon_a \left(\frac{1}{4}-\delta_{\mu 0}\right),
\end{equation}
which has no divergence, but has the same curl as the original
``source'' fields.  As promised, explicit monopole degrees of freedom
have been introduced.  A monopole number operator $N_r$ is slaved (by a
dual Gauss' law constraint) to the ${\vec B}$ field,
\begin{equation}
  \label{eq:Nslave}
  \divo {\vec B} = 2 \pi N_r.
\end{equation}
It is conjugate to the dual phase $\theta_r$, such that
\be\label{defect_commutator}
\Big[
\theta_r , \, N_{r'} 
\Big] = +i \delta_{r,r'}\;.
\ee
Eq.~(\ref{eq:Nslave}) is another $U(1)$ gauge constraint, so it is not
surprising that the monopole hopping term $w$ respects a dual ({\sl
  non}-compact) gauge symmetry.  

Although the monopole number $N_r$ (which can be both positive or
negative reflecting the two signs of flux emanating from a monopole)
appears nowhere explicitly in Eq.~(\ref{gaugedH}), it is implicit
through the constraint of Eq.~(\ref{eq:Nslave}).  In the Coulomb phase
for large $K$, therefore, monopoles are energetically costly (though
their energy is finite, as is easily verified by integrating the
associated $B^2$ energy density), with a gap of $O(K)$.  Through the
$w$ term, however, monopoles do not reside in localized states with
$N_r=\pm 1$, but instead in superpositions of such states, with only
$\sum_r N_r = \pm 1$.  As $K$ is decreased, the monopole energy gap
decreases, and at some point it will reach zero.  This point
corresponds to the confinement transition discussed in the previous section.

\section{Formalism -- Monopole defects on the diamond lattice}
\label{sec:action}

\subsection{Ground state manifold}

To understand the confinement transition, we must understand the
nature of the lowest energy monopole and anti-monopole states, which
condense at the transition.  They are equivalent by
$\vec{B}\rightarrow -\vec{B}$ symmetry, so it is sufficient to study
just the monopole states.  The ultimate field theory will consist of a
relativistic field for each member of the monopole multiplet, since
the relativistic description includes particles and antiparticles
(here antimonopoles) on equal footing.  We will apply for the most
part a (dual) mean-field approach, taking $\curl \alpha =
\overline{e}$ in \eqref{gaugedH}, and neglecting fluctuations of
$\alpha$ around this value.  This is sufficient to analyze the
spectrum of ordered phases near the $U(1)$ quantum liquid.
Fluctuations will be restored later in Sec.~\ref{sec:rg-analysis}.

We may thus consider the manifold of states with one monopole, i.e.
$N_r=1$ on one and only one site of the dual lattice, and $N_r=0$ on
all other sites.  Through the $w$ term in \eqref{gaugedH}, 
the wavefunction of the
monopole delocalizes, and is described by a tight-binding model, which
we may write as 
\be\label{H1} {\mathcal H_{tb}} = - w \sum_{ \langle r , r'
  \rangle } \{ {\psi^\dag}(r') \psi(r) e^{-i \alpha_{r,r'}} + h.c.  \}
\; , 
\ee 
where ${ \langle r , r' \rangle }$ denotes a summation over
nearest neighbors on the diamond lattice. Here $\psi_r^\dagger,
\psi_r^{\vphantom\dagger}$ are creation/annihilation operators for the
monopole.  Note we have absorbed a factor of $2\pi$ into the vector
potential relative to \eqref{gaugedH} to make our notation more
conventional.  By our mean-field assumption, $\alpha_{r,r'}$ is a
c-number vector potential carrying a ``flux'' (actually electric flux)
given by $2\pi\overline{e}$.  Since it appears only in a periodic
exponential, the form in Eq.~(\ref{eq:ebar}) is equivalent to a flux
of $2\pi/4=\pi/2$ through each dual plaquette.

\begin{figure}
        \centering
                \ifig[height=2.5in]{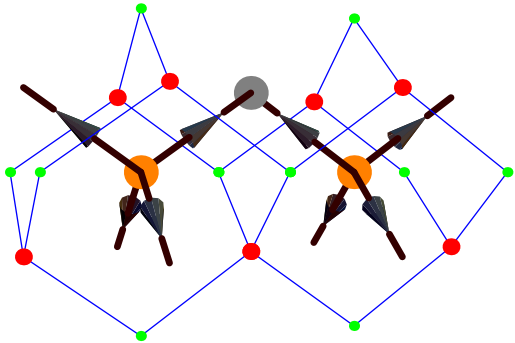}
                \caption{(Color online.) Alternating charge distribution emits
                `magnetic' field lines through the faces of a diamond lattice cell.}
        \label{fig:field_lines}
\end{figure}

This can be understood as follows.  The original compact QED theory
had a staggered background charge of $\epsilon_a$ on each direct
lattice site.  The monopoles, as magnetic charges, see this in the
same way electric charges would see a staggered lattice of magnetic
monopoles and anti-monopoles.  These `monopoles' and `anti-monopoles'
(we use single quotes to denote the dual view, since these are
actually the background gauge charges) are distributed in an
alternating fashion at the center of each cell of the dual diamond
lattice. The neighboring cells to a cell containing a `monopole' all
contain `anti-monopoles'.  Each `monopole' has a ``charge'' of 2$\pi$.
Since all lattice directions are equivalent, the `magnetic' flux going
out of each face of the cell must be the same, as illustrated in
Fig.~\ref{fig:field_lines}. The structure of the diamond lattice is
made of cells where each cell has 4 faces, as opposed to the cubic
lattice which has 6 faces for each of its cells (cubes). Thus we
conclude that each face in the diamond lattice has a flux of $
\frac{2\,\pi }{4} = \frac{\pi }{2}$ going through it in the direction
from a `monopole' cell to an `anti-monopole' cell.  The $\psi_r$
monopole particle thus experiences Aharonov-Bohm fluxes of precisely
this sort as it moves through the lattice.

It proves convenient to describe the links of the diamond lattice by
$(r,r') = (a,\mu)$ where $a \in \u$ denote the sites of the $\u$
sublattice, and $ \mu \in \{0,1,2,3\} $ enumerate the 4 links
emanating from each $\u$ site.  Furthermore, we can enumerate the $\u$
sublattice sites by $ {\vec r}=\sum_{j=1}^{3} n_j {\vec a}_j $ where
${\vec a}_j$ [${\vec a}_1=\frac{a}{2}(\hat y +\hat z), {\vec
a}_2=\frac{a}{2}(\hat x +\hat z), {\vec a}_3=\frac{a}{2}(\hat x +\hat
y)$] are the primitive Bravais lattice vectors of the FCC lattice, and
$n_j$ span the integer numbers. We refer to this coordinate system as
``index'' space.

We shall now focus our attention on finding the ground state manifold 
of this Hamiltonian. First we must find an appropriate choice of the 
vector potential giving the desired flux pattern through the faces inside 
the lattice. To this end, the ``index space'' notation proves 
particularly useful. One such possible vector potential is
\be
\begin{split} &
\alpha_0(\vec n) = 0 \; ,
\\ &
{\vec \alpha}(\vec n) = 
\left(
\alpha_1(\vec n) ,
\alpha_2(\vec n) ,
\alpha_3(\vec n)
\right)
\equiv
{\vec \epsilon}
\left( {\vec Q} \cdot {\vec n} \right) \; ,
\end{split}
\ee 
where $ \vec Q = \frac{\pi}{2} \left(1,0,-1\right) $ and 
$ {\vec \epsilon} = \left(1,1,2\right) $. 
%The chosen form is linear in the ``index space'' 
%position.

\begin{figure}[!hbp]
\begin{center}
%\vskip+6mm
\ifig[height=2.5in]{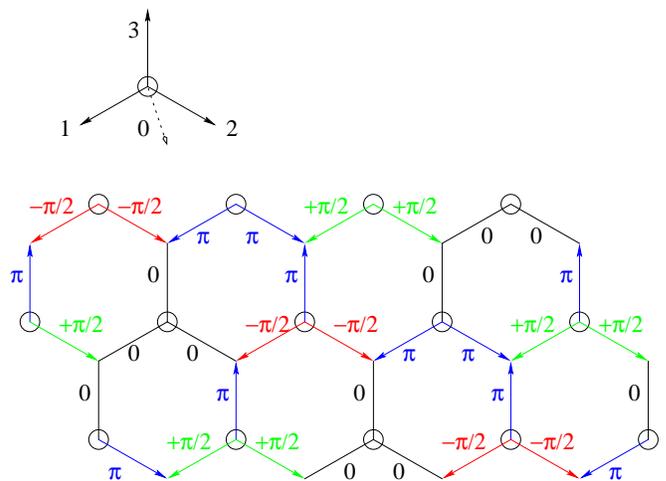}
\caption{(Color online.) Projected diamond plane view of vector potential pattern -- links between the
honeycomb planes have $\alpha_0 = 0$. All links with the same vector potential value are in the 
same color.}
%\vskip-4mm
\label{fig:alpha_pattern}
\end{center}
\end{figure}

We proceed to diagonalize the hopping term. General eigenstates cannot
be found analytically, however minimum energy eigenstates can. We find
8 ground state eigenmodes, denoted $\Phi_{\nu}$ and ${\overline
  \Phi}_{\nu}$ where the indices run through $\nu = 0,1,2,3$. The
details of the 8 eigenmodes are are left for
Appendix~\ref{app:GS_modes}.

\subsection{Symmetries} 

The 8 k-space minima can be related by the symmetries of the Hamiltonian. Each
one of the symmetries is represented by an operator that commutes with
the Hamiltonian. However, some of the symmetry operators do \emph{not}
commute with each other. If we choose to represent the Hilbert 
subspace of the minimum
energy states of the Hamiltonian using a basis of eigenstates that are
also eigenstates of a symmetry operator $ {\hat S}_1 $, then for a
different operator $ {\hat S}_2 $ with which it does not commute $
\lbrack {\hat S}_1 , {\hat S}_2 \rbrack \neq 0 $ the basis states will
\emph{not} be eigenstates of $ {\hat S}_2 $.  Thus by acting with
$\hat{S}_2$ on the eigenstates of $\hat{S}_1$, one may generate
another linearly independent eigenstate of the Hamiltonian with the
same energy.  Such a structure therefore constrains the minimum
size of ground state multiplets.  This idea should be familiar from the
representation theory of $SU(2)$ in standard quantum mechanics.

While the symmetry group of the diamond lattice is the space group Fd${\bar
  3}$m~\cite{Bilbao}, the dual Hamiltonian itself transforms under a
\emph{projective symmetry group} (PSG), since a vector potential $\alpha$
appears explicitly. The PSG differs from the space symmetry group by a
specific gauge transformation accompanying every space symmetry
operation.

We can describe the entire symmetry group using a reduced set of
operators -- generators -- from which any symmetry operation can be
constructed as a product of members in the reduced set. One can find a
minimal such reduced set. We shall consider the minimal set of the
point symmetry group, as well as the 3 primitive translations. We
leave the complete description of the PSG to Appendix~\ref{app:PSG},
and only state here that the reduced set comprises
\begin{itemize}
\item 3 (FCC) translations ${\mathit t_j}$,
\item a reflection symmetry ${\mathit i}$,
\item an inversion symmetry ${\mathit p}$,
\item a 2-fold rotation symmetry ${\mathit r_2}$,
\item a 3-fold  rotation symmetry ${\mathit r_3}$.
\end{itemize} 
Our convention is to denote symmetry operations in the space symmetry group by lowercase letters, and 
the corresponding PSG operations by uppercase letters. The space group
obeys some group algebra, which is just a multiplication table.  It
can be constructed by imposing a set of algebraic rules on the
generators alone, of the form
\be\label{rules1}
{\hat s}_1 \ldots {\hat s}_n = 1 \;,
\ee
where ${\hat s}_i$ are generators.
The details of these relations are left to Appendix~\ref{app:algebra}.

To obtain the PSG we must add gauge transformations to each of these symmetry operations.
Consider a symmetry operation $\hat S$ with the following action on the 
lattice sites in real space coordinates
${\hat s} : \, r \longrightarrow r' $. Now add a gauge transformation to accompany the lattice site transformation
\be
{\hat S} : \, {\psi}(r) \longrightarrow   
{\psi}(r')
\, e^{-i \Lambda (r) } \; .
\ee
Let us examine what this transformation does to a generic hopping term in the Hamiltonian
\be \label{Gauge_finding_logic}
\bs &
{\hat S} : \, {\psi}^{\dag} (r_2) \, {\psi}(r_1) 
\, e^{-i \alpha_{r_1,r_2}}
\longrightarrow   \\ &
{\psi}^{\dag} (r_2 ') \, {\psi}(r_1 ')
\, e^{+i \left( \Lambda (r_2) - \Lambda (r_1) 
- \alpha_{r_1,r_2}
\right) } \; .
\end{split}
\ee
We require that the Hamiltonian be invariant under this transformation, and so we must have
\be
e^{+i \left( \Lambda (r_2) - \Lambda (r_1) 
- \alpha_{r_1,r_2}
\right) } = 
e^{-i \alpha_{r_1 ' ,r_2 '} }
\; ,
\ee
or put more simply
\be
\Lambda (r_2) - \Lambda (r_1) 
- \alpha_{r_1,r_2} = 
- \alpha_{r_1 ' ,r_2 '} \pmod{2\pi}
\; .
\ee
Using this procedure we can find the appropriate gauge transformations for each of the 
symmetry operations in our reduced set. We leave the details to Appendix~\ref{app:PSG}.

The PSG has a modified group algebra.  The relations among the
generators differ only slightly from those in the space group, which
are given in \eqref{lattice_algebra}.  Consider any succession of
symmetry operations that takes every site on the lattice back to
itself. The same succession of the ``gauged'' operations can therefore
only perform a gauge transformation on the lattice model. Since all
the gauged symmetry operations leave the Hamiltonian invariant, and
the various terms in the Hamiltonian depend explicitly on the vector
potential, any gauge transformation that is not uniform (a global
gauge transformation) will modify the vector potential by introducing
non-zero gradients in the gauge transformation. We conclude,
therefore, that any succession of symmetry operations that leave the
lattice sites in place can only undergo an additional {\it{global}}
gauge transformation when those operations are gauged. A rule in the
algebra \eqref{rules1} is in general modified into
\be\label{rules2} {\hat S}_1 \ldots {\hat S}_n = e^{+i \theta} \;, \ee
where $\theta$ is some angle.  The PSG algebra differs from the space
symmetry group algebra \eqref{lattice_algebra} only in the following
rules \be\label{lattice_algebra2} {\mathit t}_{j+1}^{\phantom{2}} \cdot {\mathit
  t}_{j}^{\phantom{2}} \cdot {{\mathit t}_{j+1}}^{-1} \cdot {{\mathit t}_{j}}^{-1} =
1 \; , \ee in the space algebra, and \be\label{lattice_algebra3}
\T{j+1}^{\phantom{2}} \cdot \T{j}^{\phantom{2}} \cdot {\T{j+1}}^{-1} \cdot {\T{j}}^{-1} = -i \; ,
\ee in the PSG algebra.

The 8 ground state modes we found span the ground state manifold, and so any linear combination of these 
states is also a ground state,
\be\label{gs_manifold}
\Psi = 
\sum_{\nu=0}^{3} \Big( 
\phi_{\nu ,\u} \, \Phi_{\nu} + 
\phi_{\nu ,\d} \, {\overline \Phi}_{\nu}
\Big)
\; ,
\ee
where $\phi_{\nu ,a}$ now denote the complex amplitudes of the 8 eigenmodes.
Given the symmetry operations in momentum space, we find the ground state manifold is closed and
completely connected -- no disconnected subsets of the manifold exist.

We now assume that these 8 fields are slowly varying with position on
the lattice, so we can treat them using a continuum limit. However,
these 8 fields will still be required to respect the symmetry of the
underlying lattice.

The 8 slowly varying fields $\phi_{\nu ,a}$ transform under a particular
8-dimensional irreducible representation (irrep) of the PSG. In fact, we
prove in appendix~\ref{app:8dim_proof} that this is the minimum
dimension of a representation of the PSG.  It is convenient to change
the field basis to \be\label{field_trans}
\begin{split}
  {{{{\phi }_{0\u}}}= {\frac{{{\zeta }_0} + {{\zeta
          }_1}}{{\sqrt{2}}}}} \; , \qquad {{{{\phi }_{2\u}}}=
    {\frac{{{\zeta }_0} - {{\zeta }_1}}{{\sqrt{2}}}}} \; ,
  \\
  {{{{\phi }_{1\u}}}= {\frac{{{\zeta }_2} + {{\zeta
          }_3}}{{\sqrt{2}}}}} \; , \qquad {{{{\phi }_{3\u}}}=
    {\frac{{{\zeta }_2} - {{\zeta }_3}}{{\sqrt{2}}}}} \; ,
  \\
  {{{{\phi }_{0\d}}}= {\frac{{{\xi }_0} + {{\xi }_1}}{{\sqrt{2}}}}}
 ,  \qquad \;  {{{{\phi }_{2\d}}}= {\frac{{{\xi }_0} - {{\xi
          }_1}}{{\sqrt{2}}}}} \; ,
  \\
  {{{{\phi }_{1\d}}}= {\frac{{{\xi }_2} + {{\xi }_3}}{{\sqrt{2}}}}} \;
  , \qquad {{{{\phi }_{3\d}}}= {\frac{{{\xi }_2} - {{\xi
          }_3}}{{\sqrt{2}}}}} \; .
\end{split}
\ee
This basis realizes a ``permutative representation'' of the PSG, in
the nomenclature of Ref.\cite{Balents_1:prb05}.  That is, the symmetry
operations of the PSG act on these fields by a combination of
permutations and simple diagonal phase rotations.  See
Eqs.~(\ref{eq:prep})-\eqref{eq:prep_f}.  Because of this structure, 
the action takes a particularly simple form in this basis.

\subsection{Effective low energy action}

Our goal is to describe a low-temperature condensate phase of the
monopole defects. Limiting the discussion to zero temperature, we set
out to construct a phenomenological Landau-Ginzburg (LG) action to
access the condensate phase. We concentrate on the ground state
manifold of the monopole defects, ignoring any higher energy modes,
and construct a low energy effective continuum action in the 8 field
components of \eqref{gs_manifold}.  The various terms allowed in the
action must be invariant under the PSG in the 8-dimensional representation
\eqref{eq:prep}-\eqref{eq:prep_f}. As the effective action will live in a 3+1 dimensional
spacetime, we seek terms only up to quartic order in the field
operators -- any higher order terms will be irrelevant in the
renormalization group sense.

To quadratic order only one invariant exists,
\be\label{quad_inv}
\Theta_1 = \sum_j \left( |\zeta_j|^2 +|\xi_j|^2 \right)
\; ,
\ee
a typical mass term, where the indices enumerate $j=0,1,2,3$.

At quartic order we find 4 invariants \be
\begin{split} \label{eq:kappa_text}&
(\Theta_1)^2 = (|\vec{\zeta}|^2 +|\vec{\xi}|^2)^2
\; ,
\\ &
\Theta_2 =\sum_{i \neq j} |\zeta_j|^{2} |\zeta_i|^{2} 
+ \sum_{i \neq j} |\xi_j|^{2} |\xi_i|^{2} 
\; ,
\\ &
\Theta_3 = |\vec{\zeta}|^2 |\vec{\xi}|^2 
\; ,
\\ &
\Theta_4 = \sum_{ijkl} \kappa_{i j}^{\phantom{i j} k l} \zeta_k^{\phantom{*}} \xi_l^{\phantom{*}} \zeta_{i}^* \xi_{j}^* 
\; ,
\end{split}
\ee
where the sums over indices are always over $0,1,2,3$ unless otherwise stated, and the vector notation
implies a 4-component vector. We shall not specify the $\kappa$
tensor explicitly, as it is long and complicated. 
We leave it to Appendix~\ref{app:kappa_tensor}.

The new set of independent terms obeys numerous continuous symmetries
in a rather transparent manner (for brevity we do not specify the
discrete symmetries of these terms):
\begin{itemize}
\item $\Theta_1$ is invariant under a full $U(8)$ symmetry, as it is
  just the norm of an $8$ dimensional complex coordinate vector.
\item $\Theta_2$ is invariant under a $[U(1)]^8$ symmetry, since it
  depends only upon the magnitudes of the field components in the new
  basis.  We may change freely and independently the phase of each
  field component.
\item $\Theta_3$ is invariant under a $U(4) \times U(4)$ symmetry,
  because it involves only the norm of two $4$ dimensional complex
  vectors.
  \item $\Theta_4$ obeys a $[U(1)]^2$ symmetry group corresponding to
  the following transformation \be\label{symm_5}
  \begin{split} \forall j \quad &
    \zeta_j \rightarrow e^{+i \delta}\zeta_j
    \; ,
    \\ &
    \xi_j \rightarrow e^{+i \lambda}\xi_j
    \; .
  \end{split}
  \ee
  \end{itemize}
The action as a whole is invariant under a $[U(1)]^2$ symmetry, governed
by the $\Theta_4$ invariant.

The only microscopic symmetry of the action is the dual $U(1)$ gauge
invariance, i.e. identical phase rotations of all field components.
This changes just the overall phase of the monopole wavefunction, and
has no physical consequence (it is a true gauge). The other
``staggered'' $U(1)$ rotation (with $\delta=-\lambda$ in
Eq.~(\ref{symm_5})) is accidental, occurring only in the proximity of
the critical point at which truncation to the quartic action is a good
first approximation.  It will be broken if we include sufficiently
high order terms in the action.  Our investigation concluded that the
staggered $U(1)$ symmetry persists at 6th order, but ultimately breaks
down to a discrete symmetry ($Z_4$) at 8th order. The remaining discrete symmetry $Z_4$ 
can be identified as part of the lattice PSG.

Finally, the most general low energy effective action up to quartic order is
\be\label{action3}
\begin{split} &
{\mathcal S} = 
\int_{\bf r, \tau}
\Big\{
\left| \left( \partial_{\mu} - i \alpha_{\mu} \right) \zeta_j \right|^{2}
+ \left| \left( \partial_{\mu} - i \alpha_{\mu} \right) \xi_j \right|^{2}
+ \tilde \gamma  \Theta_1
\\  &
+ \gamma_1  \Theta_1^2
+ \gamma_2 \Theta_2 + \gamma_3 \Theta_3 + \gamma_4 \Theta_4
+ \frac{1}{2 e^2} F^2
\Big\}
\; ,
\end{split}
\ee where a sum over $j$, as well as integration over spacetime are
implied, and $\frac{1}{2 e^2} F^2$ is the Maxwell term in the action, with
$F_{\mu\nu} = \partial_\mu \alpha_\nu - \partial_\nu \alpha_\mu$. Here $\alpha$ is the 
continuum version of the vector potential $\alpha$ appearing in Eq.~\eqref{H1}. The
$\gamma_j$ are phenomenological couplings undetermined in our theory. 

Higher order terms can be ignored for a renormalization group (RG)
analysis (in $3+1$ dimensions, they are irrelevant in the RG sense,
and can be treated perturbatively when necessary), but must be taken
into account in a mean field analysis to which we now turn.

\section{Mean field theory}
\label{sec:mean-field}

We now turn to analyze our effective action \eqref{action3}.  Using
mean-field theory (MFT) we can find the various phases of this action.
Replacing the fields with their average values, we obtain a Lagrangian
density ${\cal L}$ that must be minimized with respect to the field averages.

At first we ignore the $\Theta_4$ term, since it has the lowest
symmetry. We shall examine only the ordered phases which occur for weak
$\gamma_4$ (and higher order coefficients $\gamma_5,\gamma_6$ when
necessary -- see below) for simplicity.  Scaling all the couplings to
that of ${\Theta_1}^2$ we get \be\begin{split} & M^2 = -
  \frac{\tilde \gamma}{\gamma_1} \; , \\ & \beta_1 = \frac{\gamma_2}{\gamma_1} \; ,
  \\ & \beta_2 = \frac{\gamma_3}{\gamma_1} + 2 \; , \\ & \beta_3 =
  \frac{\gamma_4}{\gamma_1} \; .
\end{split}
\ee 
The mean-field Lagrangian density with $\beta_3 = \gamma_4=0$ is
\be
\begin{split}
{\cal L}/\gamma_1 =  & \left (
- M^2 |\vec \zeta|^2 + (|\vec \zeta|^2)^2 + \beta_1 \sum_{i \neq j} |\zeta_j|^{2} |\zeta_i|^{2}
\right) + (\zeta \rightarrow \xi)
\\ &
+ \beta_2 |\vec{\zeta}|^2 |\vec{\xi}|^2
\; .
\end{split}
\ee

\begin{figure}%[!hbp]
\begin{center}
%\vskip+6mm
\ifig[height=2.0in]{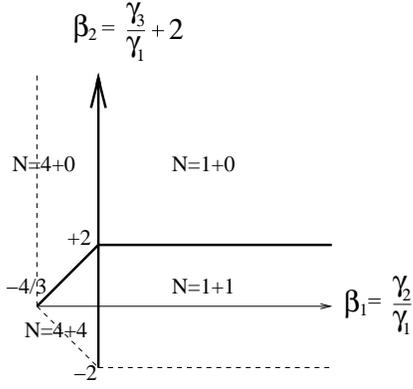}
\caption{(Color online) Mean field theory phase diagram of simplified free energy, $\beta_3 = 0$. 
The phases exist only within the bounds of the dashed lines. Different phases are delimited by 
continuous lines.}
%\vskip-4mm
\label{fig:MFT}
\end{center}
\end{figure}

We first find the mean-field minima with $M^2 > 0$ in the $\beta_{1,2}$ plane.
The resulting phase diagram is illustrated in Fig.~\ref{fig:MFT}.  We
label the ``phases'' by the number of non-zero components of
$\vec\xi$ and $\vec\zeta$, ordered from largest to smallest, with the
notation $N_1 + N_2$.  The ``phases'' are:
\begin{itemize}
\item $ \beta_1 > 0 \, , \beta_2 > 2$, \newline
In this region, only one field component condenses, so this region is
labeled $1+0$. 
\item $ \beta_1 > 0 \, , -2 < \beta_2 < 2 $, \newline 
Here one field component of each set ($\zeta_i, \xi_j$) condenses, so
this is a $1+1$ region.
\item $ \beta_1 < 0 \, , \beta_2 > 2\left(1+\beta_1 \frac{3}{4}\right)
  $, \newline  
  Here all the field components of one of the 2 sets condense, denoted
  $4+0$.
\item $ \beta_1 > 0 \, , -2\left(1+\beta_1 \frac{3}{4}\right) < 
\beta_2 < 2\left(1+\beta_1 \frac{3}{4}\right) $, \newline
Here all the 8 field components are non-zero, so it is a $4+4$ region.
\end{itemize}
It should be pointed out that specifying the number of non-zero
components of the {\sl complex} fields does not necessarily specify a
unique charge-ordered {\sl phase} of the problem.  Physically, these
charge-ordered states break only discrete space group symmetries, and
so are expected to lead to only a discrete degeneracy of ground
states.  In our dual gauge formulation, each physically distinct state
is represented by a ``cycle'' in the monopole field space, along which
the phase of all the fields is varied together by a uniform angle
which can go from $0$ to $2\pi$.  This variation has no physical
significance since this is a gauge symmetry.  Other continuous phase
freedoms of the mean field solutions, which do not vary the phases of
all components identically, are however not a pure gauge freedom. Such a
(artificial) freedom can only be a
result of truncating the dual action to quartic order.

In fact, only the $1+0$ phase lacks any such ``emergent'' phase
freedom.  In particular, the different $1+0$ states obtained by
choosing each of the 8 possible components non-zero are physically
distinct, and the phase of this non-vanishing field is pure gauge,
with no physical significance.  Thus, the $1+0$ states are eight-fold
degenerate.  The $\beta_3$ term vanishes for such states and does not
change this conclusion. These states exhibit an enlarged unit cell, 
comprising 4 unit cells of the underlying lattice, and
containing 16 pyrochlore lattice sites (In fact, we will find that all our mean-field 
states have an enlarged unit cell. The amount by which the unit cell
is enlarged can be computed by explicitly checking how a given mean-field
state transforms under an arbitary translation using the PSG.)
The Bravais lattice formed by these enlarged unit cells is simple cubic (SC).
It is noteworthy that this phase is the R state found in 
Ref.~\onlinecite{Bergman:prl05}.

The other $1+1$, $4+0$, and $4+4$ solutions retain at least some
physically meaningful phase freedom even when $\beta_3$ is included in
the action.  Consider first the $4+0$ case.  For such configurations,
the $\beta_3$ term vanishes, so the phases of all four non-zero fields
remain free at quartic order.  Only the overall phase of the four
fields is gauge, and hence will remain undetermined -- and unphysical
-- to all orders.  To fix the remaining three phases, we must consider
higher order terms.  In this case, 6$^{th}$ order is sufficient, and
there is one term which resolves the continuous degeneracy:
\begin{eqnarray}
  \label{eq:L6}
  {\cal L}_6 & = & \gamma_5 {\rm Re}\Big\{ e^{-i\pi/4}\big[
  (\zeta_0^*)^3 \zeta_1^{\vphantom*} \zeta^{\vphantom*}_2
  \zeta^{\vphantom*}_3 -(\zeta_1^*)^3 
  \zeta^{\vphantom*}_0 \zeta^{\vphantom*}_2 \zeta^{\vphantom*}_3
  \nonumber \\ & & -(\zeta_2^*)^3 \zeta^{\vphantom*}_0
  \zeta^{\vphantom*}_1 
  \zeta^{\vphantom*}_3 + (\zeta_3^*)^3 \zeta^{\vphantom*}_0 \zeta^{\vphantom*}_1 \zeta^{\vphantom*}_2\big] -
  (\vec{\zeta}\rightarrow \vec{\xi})\Big\}.
\end{eqnarray}

In the $N=4+0$ phase  for $\gamma_5 > 0$, one minimum energy state is
\be
{\vec \zeta} = \left( 1 , e^{+i \frac{3\pi}{4}} , e^{+i \frac{\pi}{4}} , 1 \right).
\ee
It is part of a set of 792 total states, which can be generated from
this one by the PSG.
For $\gamma_5 < 0$,  one configuration is
\be
{\vec \zeta} = \left( 1 , e^{+i \frac{\pi}{4}} , e^{+i \frac{\pi}{4}} , 1 \right),
\ee
which is part of a set of 384 degenerate states connected by the PSG.
The enlarged unit cell comprises 16 unit cells of the original lattice, and contains 
64 pyrochlore lattice sites. The ``supercells'' are arranged in a body-centered cubic (BCC) Bravais lattice.

In the $1+1$ case, there is a single undetermined physical
phase, an opposite phase rotation of $\vec{\zeta}$ and $\vec{\xi}$.
This is a direct result of the ``staggered'' $U(1)$ symmetry of the
action at 6$^{\rm th}$ order.  Any such solution breaks this staggered
$U(1)$, and so making such a rotation gives a different solution.  The
staggered $U(1)$ symmetry is broken only at 8$^{\rm th}$ order, by 8
distinct terms.  In the $1+1$ case, only one of these terms is 
non-vanishing.  It takes the form
\begin{eqnarray}
  \label{eq:L8a}
  {\cal L}_8 & = & \gamma_6 {\rm Re}\, \Big\{ \big[(\zeta_0^*)^4
  -(\zeta_1^*)^4  -(\zeta_2^*)^4 +(\zeta_3^*)^4 \big]\nonumber \\
  & & \times \big[\xi_0^4
  -\xi_1^4 -\xi_2^4 +\xi_3^4 \big]\Big\}.
\end{eqnarray}
The nature of the phase depends upon the sign of $\gamma_6$.
In each case, there are 64 physically distinct ground states: 4
choices each for the non-zero component of $\vec\zeta,\vec\xi$, and 4
inequivalent phase minima of Eq.~(\ref{eq:L8a}).

Of the many states we mention one configuration for $\gamma_6 < 0$
\be
\zeta_0 = \xi_0 = 1 \; ,
\ee
and one configuration for $\gamma_6 > 0$\be
\zeta_0 = e^{-i \frac{\pi}{4}} \xi_0 = 1 
\; .
\ee
The enlarged unit cell comprises 16 unit cells of the original lattice, 
and contains 64 pyrcohlore lattice sites. The ``supercells'' are arranged 
in a simple hexagonal Bravais lattice,
with a ratio of $\frac{c}{a} = 4 \sqrt{1.5}$ in standard notation.

%Taking all
%possible double lattice translations in any direction, precisely half
%these moves leave the state invariant. 
%This corresponds to a unit cell with twice the number of sites of a unit
%cell of $2 \times 2 \times 2$. 

In the $4+4$ case the relative phases in each quartet are determined at
4$^{\rm th}$  
order by $\beta_3$,
and the same ``staggered'' $U(1)$ symmetry as in the $1+1$ case remains. 

For $\gamma_4 > 0$ one possible configuration is
\be
\begin{split} &
{\vec \zeta} = \left( 1 , e^{+i \frac{\pi}{4}} , e^{+i \frac{\pi}{12}} , e^{-i \frac{\pi}{6}} \right)
\; ,
\\ &
{\vec \xi} = e^{+i \lambda_4} \left( 1 , e^{+i \frac{7 \pi}{12}} , e^{+i \frac{7 \pi}{12}} , e^{-i \frac{\pi}{2}} \right)
\; ,
\end{split}
\ee
and for $\gamma_4 < 0$ a possible configuration is
\be
\begin{split} &
{\vec \zeta} = \left( 1 , e^{+i \frac{\pi}{4}} , e^{+i \frac{3 \pi}{4}} , e^{+i \frac{\pi}{2}} \right)
\; ,
\\ &
{\vec \xi} = e^{+i \lambda_4} \left( 1 , e^{-i \frac{3 \pi}{4}} , e^{-i \frac{3 \pi}{4}} , e^{-i \frac{\pi}{2}} \right)
\; .
\end{split}
\ee
Many (over 10000 in each case) other configurations are possible, and connected via the PSG to this state.

The ``staggered'' $U(1)$ symmetry is lifted by various 8$^{\rm th}$
order terms.  Depending on the various couplings of the 8$^{\rm th}$
order terms, we get that the ``staggered'' $U(1)$ breaks into the same
two $Z_4$ multiplets as the $1+1$ state does.  We shall not belabor the
details of all 8 of these terms. \eqref{eq:L8a} alone suffices to access both
multiplets.  For $\gamma_6 > 0$ \be \lambda_4 - \lambda_0 =
\frac{\pi}{4} + n \frac{\pi}{2} \ee and for $\gamma_6 < 0$ \be \lambda_4
- \lambda_0 = n \frac{\pi}{2} \; .  \ee Finally, in any one of these
cases, the enlarged unit cell comprises $4 \times 4 \times 4 = 64$ unit cells of the original lattice,
and therefore contains 256 pyrochlore lattice sites.
The ``supercells'' are arranged in a face-centered cubic (FCC) Bravais 
lattice, with the primitive lattice vectors taken 
4 times larger relative to the original FCC.

Using a direct mapping between the monopole defect density and the
spin density variations described in the next section, we can depict
the latter in Figs.~\ref{fig:dense1}-\ref{fig:dense4D}.

\section{Ordering patterns}
\label{sec:charge-order-patt}

Having found the various allowed monopole defect condensate phases in
the abstract order parameter space by algebraic considerations, we
want to identify the physical spin ($S_i^z$) ordering patterns in each
case.

The density of the monopole defects is a varying scalar density.  As it
is the only spatially varying scalar in the problem, the spin density
must have similar spatial variation. More precisely, both densities must
obey the same symmetries, \be \rho(\vec r_i) \sim \langle S^z_i\rangle \;,
\ee
where $\rho(\vec r)$ is the monopole defect density.
The monopole defect density can be found by taking the wavefunction
\eqref{gs_manifold} squared, \be \rho(\vec r) = |\Psi(\vec r)|^2 \; .
\ee The resulting density is now a function of the field values
$\phi_{\nu,a}$. By inserting the values for the different MFT phases, we
can recover the density.

In some phases, however, the symmetry breaking is not manifest in the
scalar density, but rather in the current or kinetic energy: 
\be
\begin{split}
J_{r,r'} = i \Big( \Psi^*(\vec r') \Psi(\vec r) e^{-i \alpha_{r,r'}} - c.c \Big) \; , \\
K_{r,r'} = \Big( \Psi^*(\vec r') \Psi(\vec r) e^{-i \alpha_{r,r'}} + c.c \Big)\; .
\end{split}
\ee

Both the current density and the local kinetic energy can be encoded in a complex valued vector,
\be
v_{r,r'} = \Psi^*(\vec r') \Psi(\vec r) e^{-i \alpha_{r,r'}}
\; .
\ee
The imaginary and real parts will give us (half) the current density and the local kinetic 
energy, respectively.

Each plaquette in the dual diamond lattice corresponds to a pyrochlore
lattice site at the center of the plaquette. Any monopole defect
``object'' we can define on the dual plaquettes is also defined on the
direct pyrochlore lattice sites, and encodes the symmetry of the MFT phase.
Therefore, the function must be ``similar'' to the spin density on
these sites, in the sense of giving the correct symmetry of
the latter. An appropriate function is a loop integral
(curl) of the complex current around the plaquette \be \sum_{{\vec
    r}{\vec r}'\in \hexagon}\!\!\!\!\!\!\!\!\!\!\!\circlearrowleft \;
v_{{\vec r}{\vec r'}} \sim n_{\hexagon} = n_i \; .  \ee

\begin{figure}
        \centering
                \ifig[width=3.5in]{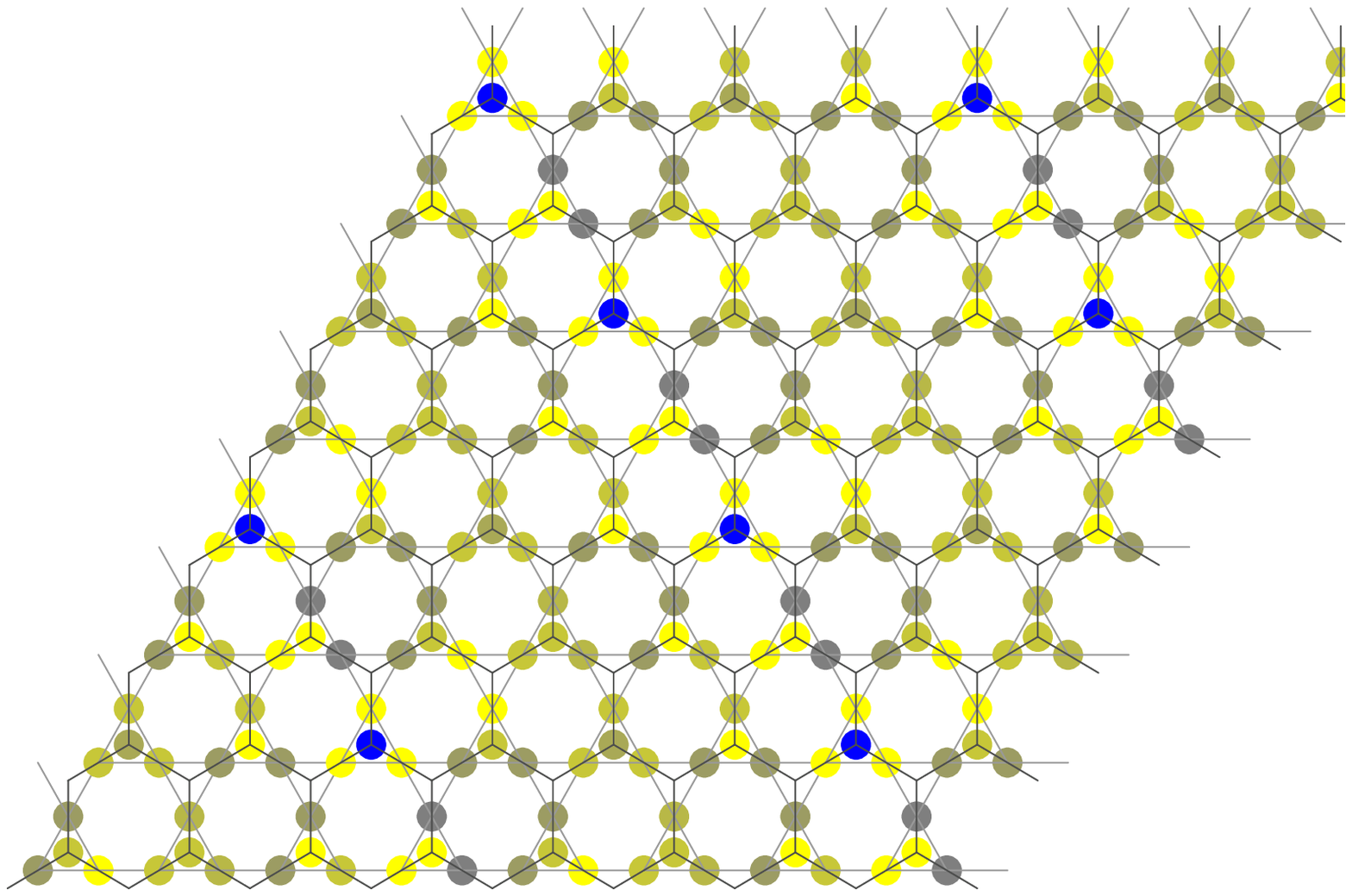}
        \caption{(Color online.) The spin density variations in the
          $1+1$ phase, for $\gamma_5 < 0 $.  One each of the four $\zeta$
          and $\xi$ fields has a non-vanishing
          expectation value.  The expectation values have identical
          magnitude.  This phase has an enlarged unit cell of $2
          \times 4 \times 2 =16$, in an simple hexagonal Bravais lattice.}
        \label{fig:dense2A}
\end{figure}

\begin{figure}
        \centering
                \ifig[width=3.5in]{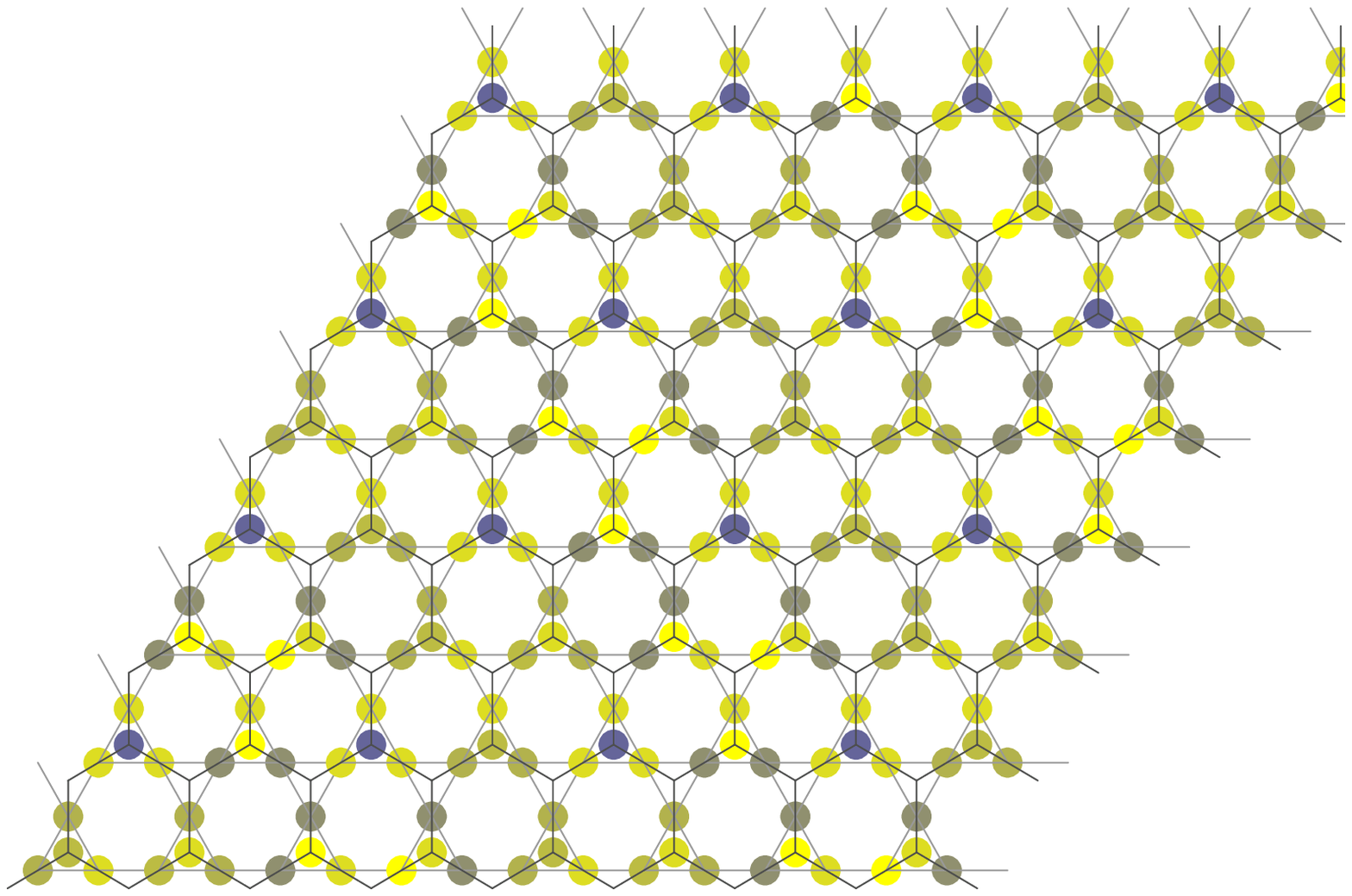}
        \caption{(Color online.) The spin density variations in the
          $1+1$ phase, for $\gamma_5 > 0 $.  One each of the four $\zeta$
          and $\xi$ fields has a non-vanishing
          expectation value.  The expectation values have identical
          magnitude.  This phase has an enlarged unit cell of $2
          \times 4 \times 2 =16$, in an simple hexagonal Bravais lattice.}
        \label{fig:dense2B}
\end{figure}

\begin{figure}
        \centering
                \ifig[width=3.5in]{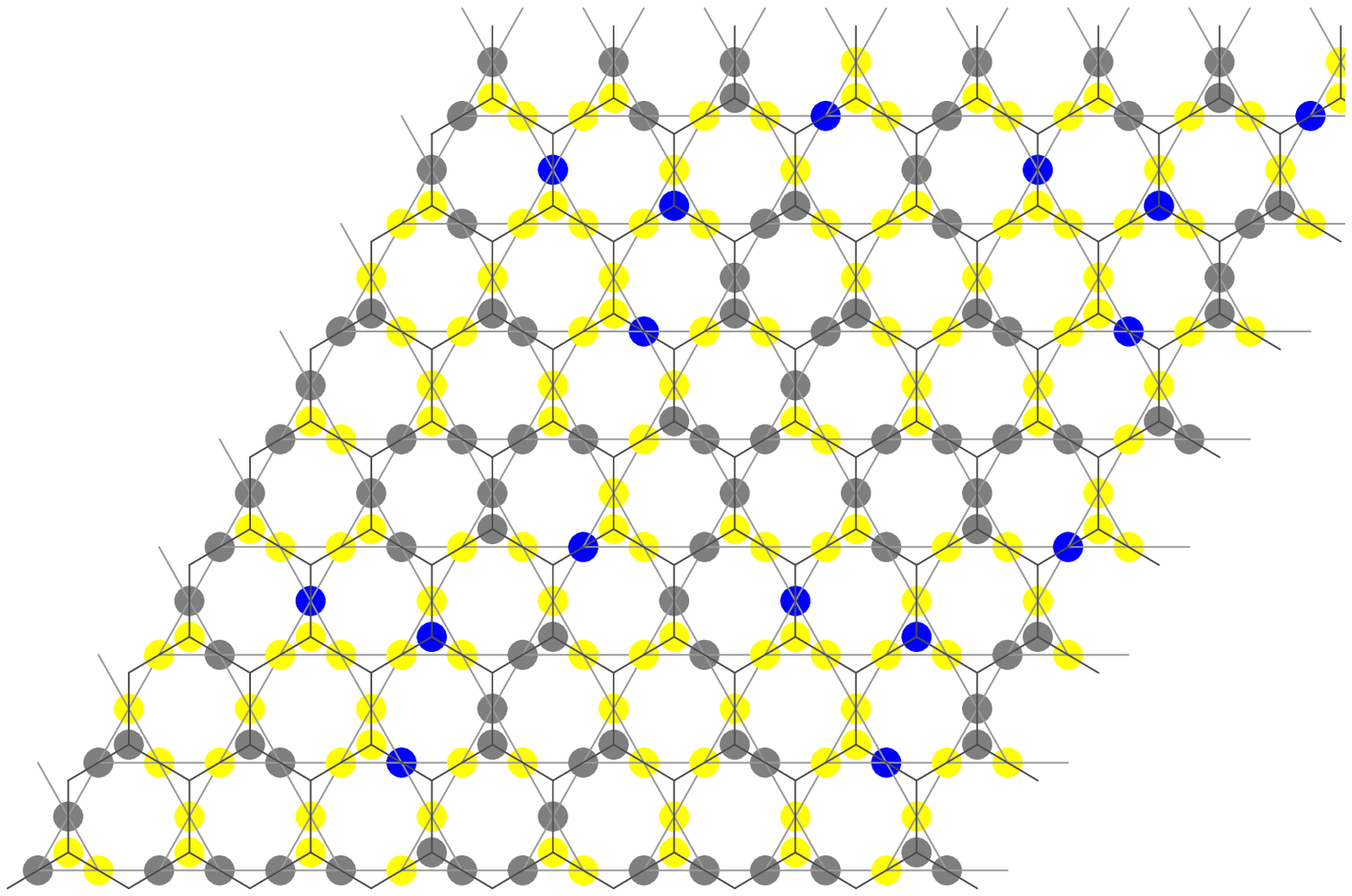}
        \caption{(Color online.) The spin density variations in the
          $4+0$ phase, for $\gamma_4 > 0 $.  The $4$ field components
          of either the $\zeta$ or $\xi$ set have a non-vanishing
          expectation value of identical magnitude.  This phase has an
          enlarged unit cell of $16$, in a BCC Bravais lattice.}
        \label{fig:dense3A}
\end{figure}

\begin{figure}
        \centering
                \ifig[width=3.5in]{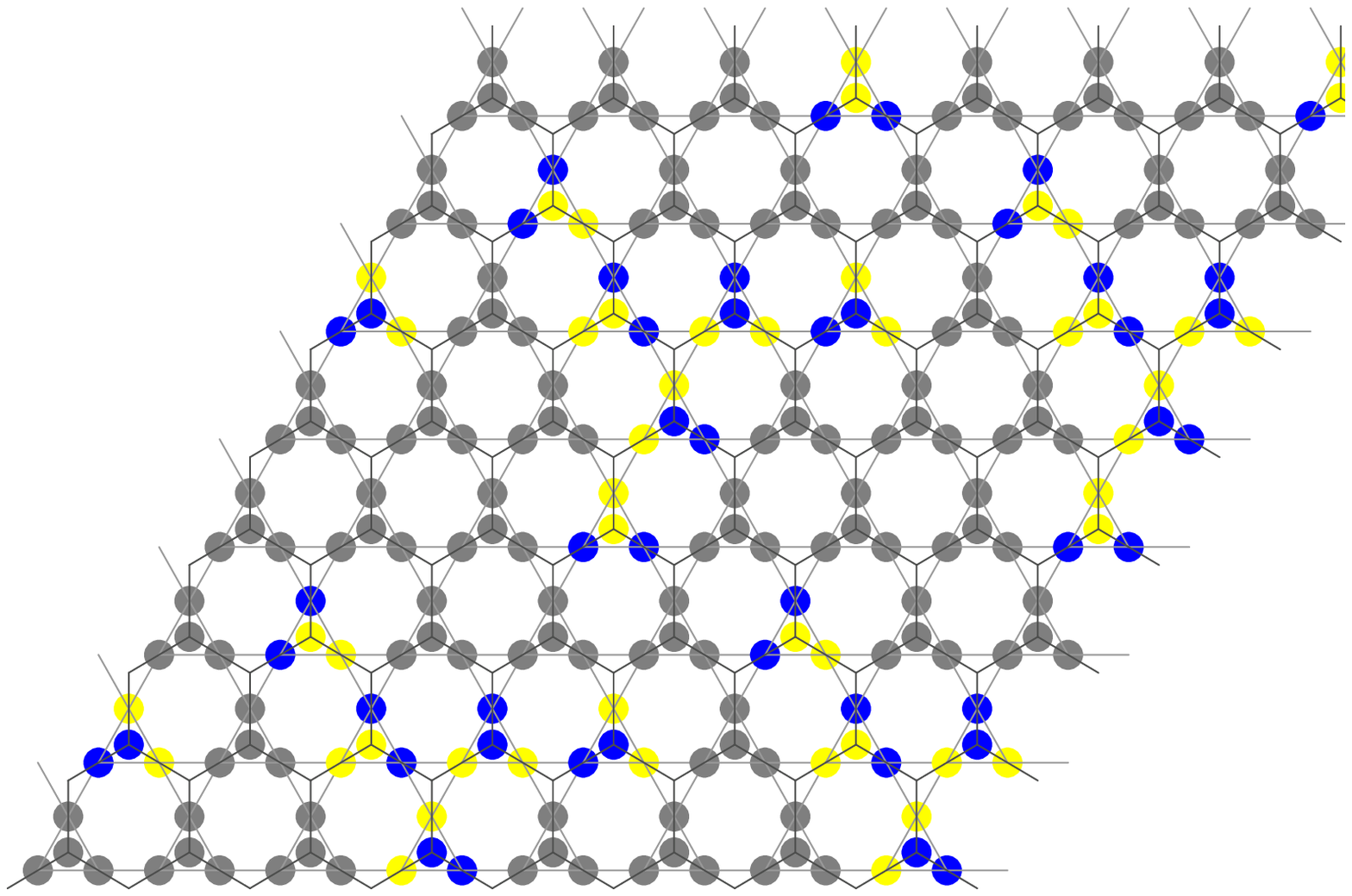}
        \caption{(Color online.) The spin density variations in the
          $4+0$ phase, for $\gamma_4 > 0 $.  The $4$ field components
          of either the $\zeta$ or $\xi$ set have a non-vanishing
          expectation value of identical magnitude.  This phase has an
          enlarged unit cell of $16$, in a BCC Bravais lattice.}
        \label{fig:dense3B}
\end{figure}

\begin{figure}
        \centering
                \ifig[width=3.5in]{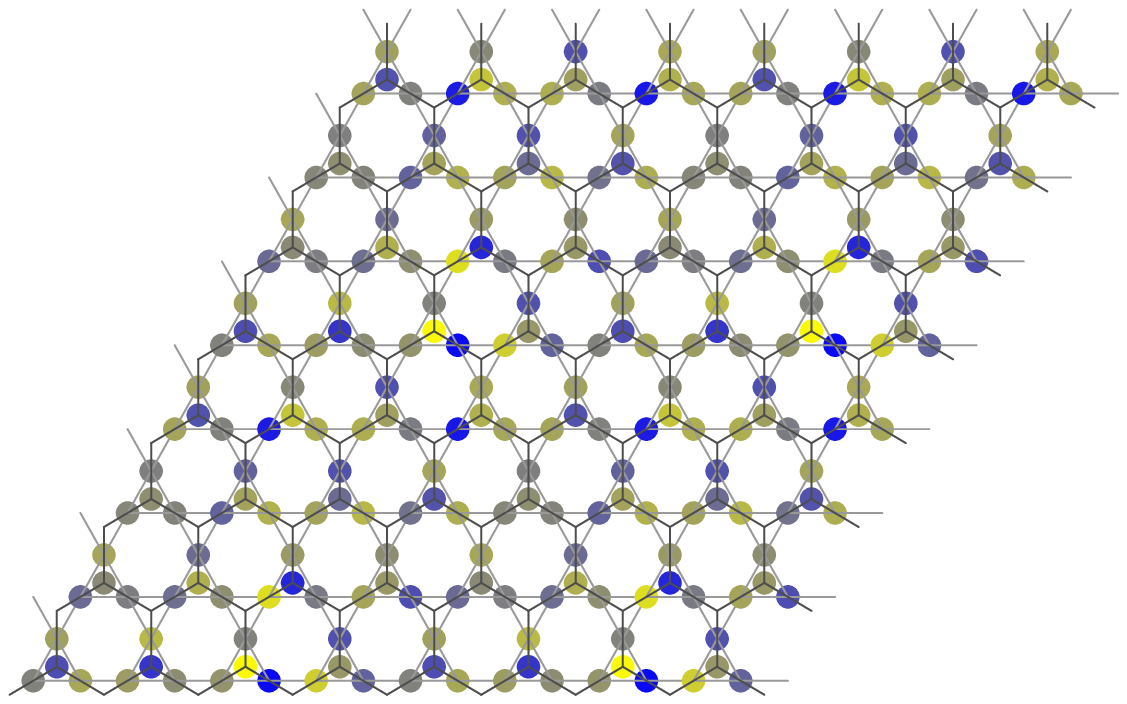}
        \caption{(Color online.) The spin density variations in the
          $4+4$ phase, for $\gamma_4 > 0 $, $\gamma_6 > 0 $.  All $8$
          field components have a non-vanishing expectation value of
          identical magnitude.  This phase has an enlarged unit cell
          of $4 \times 4 \times 4 =64$, in a FCC Bravais lattice.}
        \label{fig:dense4A}
\end{figure}

\begin{figure}
        \centering
                \ifig[width=3.5in]{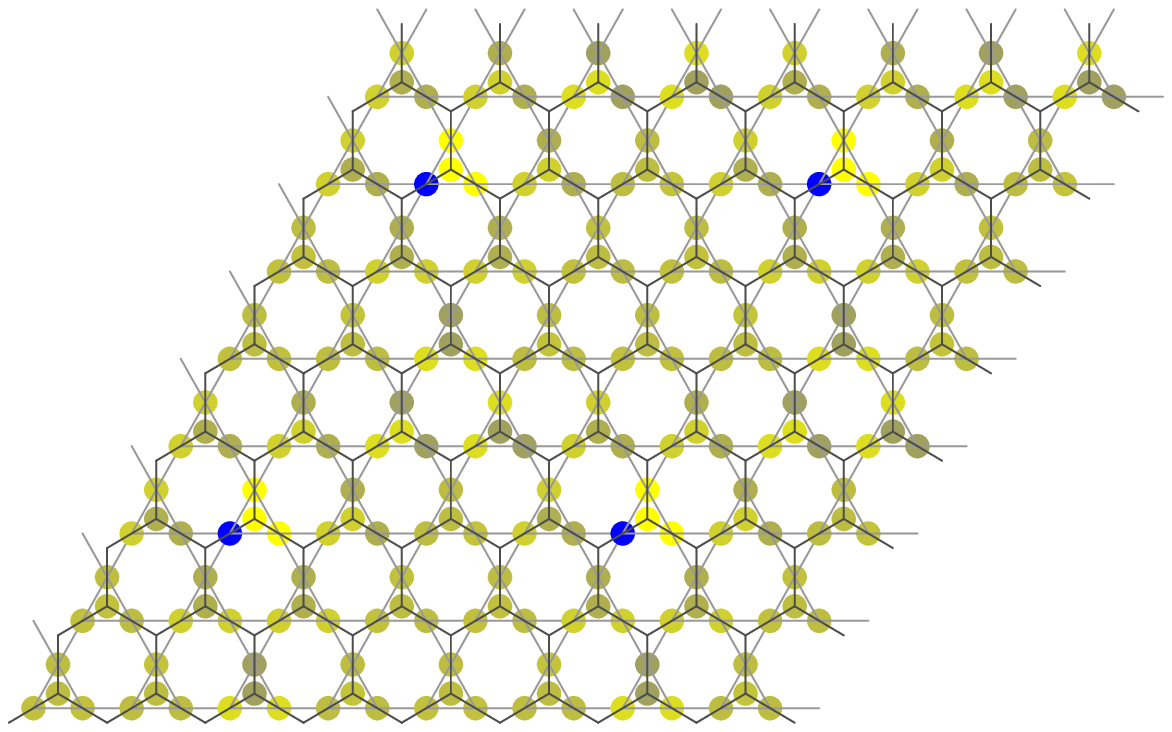}
        \caption{(Color online.) The spin density variations in the $4+4$ phase, for $\gamma_4 < 0 $, $\gamma_6 > 0 $.
        All $8$ field components have
        a non-vanishing expectation value of identical magnitude.
        This phase has an enlarged unit cell of $4 \times 4 \times 4 =64$, in a FCC Bravais lattice.}
        \label{fig:dense4B}
\end{figure}

\begin{figure}
        \centering
                \ifig[width=3.5in]{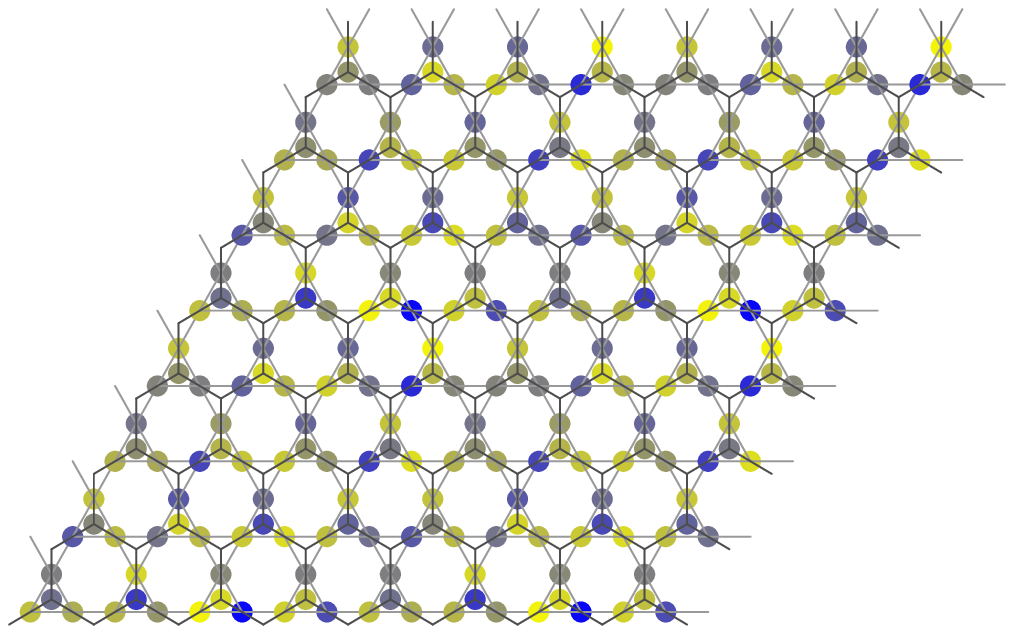}
        \caption{(Color online.) The spin density variations in the $4+4$ phase, for $\gamma_4 > 0 $, $\gamma_6 < 0 $.
        All $8$ field components have
        a non-vanishing expectation value of identical magnitude.
        This phase has an enlarged unit cell of $4 \times 4 \times 4 =64$, in a FCC Bravais lattice.}
        \label{fig:dense4C}
\end{figure}

\begin{figure}
        \centering
                \ifig[width=3.5in]{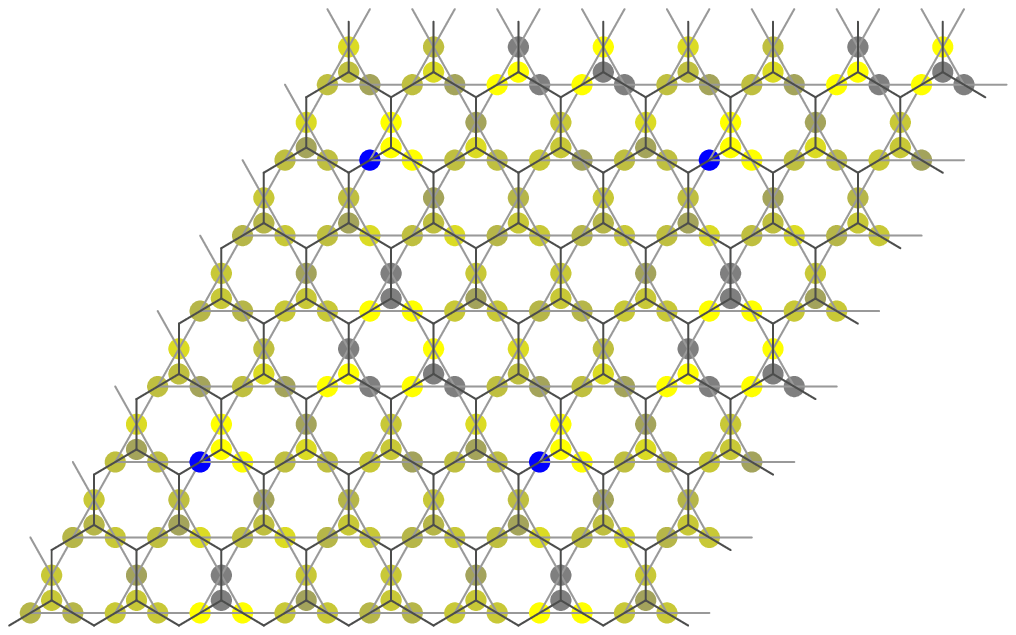}
        \caption{(Color online.) The spin density variations in the $4+4$ phase, for $\gamma_4 < 0 $, $\gamma_6 < 0 $.
        All $8$ field components have
        a non-vanishing expectation value of identical magnitude.
        This phase has an enlarged unit cell of $4 \times 4 \times 4 =64$, in a FCC Bravais lattice.}
        \label{fig:dense4D}
\end{figure}

We can formalize this argument by considering Maxwell's equations, \emph{with} magnetic monopoles
\be
\begin{split} &
\curl{\vec E} = -\frac{\partial \vec B}{\partial t} + \vec J_b \; , \\ &
\curl{\vec B} = +\frac{\partial \vec E}{\partial t} + \vec J_e \; , \\ &
\divo{\vec E} = \rho_e \; , \\ &
\divo{\vec B} = \rho_b \; .
\end{split}
\ee
where the magnetic monopole density and current are denoted with a
subscript $b$. In a static system integrating 
the first equation over some surface we get, by Stokes theorem,
\be
\oint_C \vec J_b \cdot d \vec \ell = \oint_C \curl{\vec E} \cdot d \vec \ell = \int_S \vec E \cdot d\vec A
\; .
\ee 
$ $From this last expression it is evident that the loop integral of
the monopole defect current gives the electric flux 
through that loop. The lattice version of the electric flux is a
constant plaquette ``area'' times the electric field penetrating  
perpendicular to that plaquette
\be
\sum_{{\vec r}{\vec r}'\in \hexagon}\!\!\!\!\!\!\!\!\!\!\!\circlearrowleft
\; J_{{\vec r}{\vec r'}}  = E_{a b} \sim n_i  
\; .
\ee 
In conclusion the variations in the spin density can be related to
the loop sums of the monopole defect current 
around plaquettes of the dual diamond lattice. Armed with this
knowledge we can plot a function to show the spin density 
variations. We plot the ``spin density'' so defined for each of the
phases obtained in mean field theory in
Figs.~\ref{fig:dense1}-\ref{fig:dense4D}. 

\begin{figure}
        \centering
                \ifig[height=0.8in]{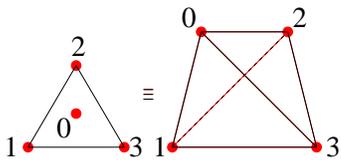}
        \caption{(Color online.) Notation for the spin density
          pictures. Triangles with a site at the center  
        represent up-pointing tetrahedra. Triangles with no site at
        the center represent down-pointing tetrahedra.} 
        \label{fig:legend}
\end{figure}

\begin{figure}
        \centering
                \ifig[width=3.2in]{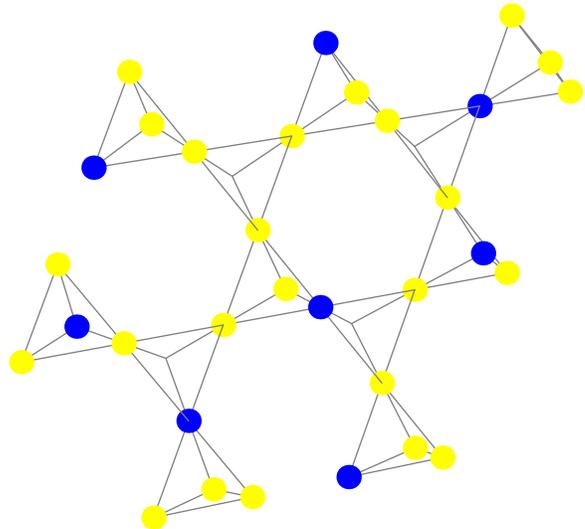}
        \caption{(Color online.) The spin density variations in the
          $1+0$ phase -- a 3D image. $1$ field component only has a
          non-vanishing expectation value. Here only minority sites in
          one kagome plane and in the triangular plane ``above'' the
          page are shown. Those in the triangular plane ``below'' the
          page are omitted to keep the image uncluttered.}
        \label{fig:dense1_3D}
\end{figure}

\begin{figure}
        \centering
                \ifig[width=3.5in]{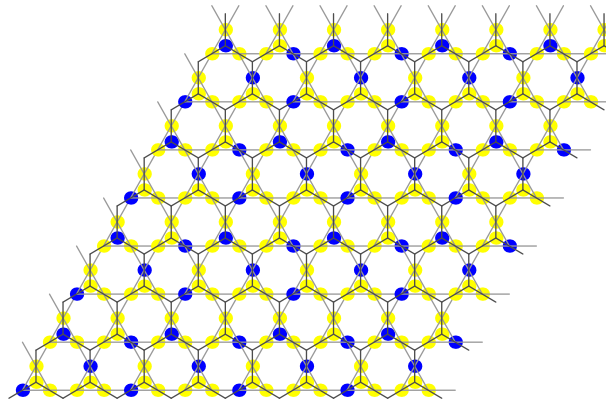}
        \caption{(Color online.) The spin density variations in the
          $1+0$ phase -- same phase as in Fig.~\ref{fig:dense1_3D} for
          comparison.  This phase has an enlarged unit cell of $2
          \times 2 \times 1 = 4$, in a simple cubic Bravais lattice.}
        \label{fig:dense1}
\end{figure}

\section{RG analysis}
\label{sec:rg-analysis}

In this section, we briefly consider the effect of fluctuations on the
mean-field critical behavior of our effective action.  Our primary
focus in this paper is {\sl not} on quantum critical phenomena, but
rather on the nature of the ordered phases {\sl close} to the $U(1)$
spin liquid state.  These results for the ordered phases are
independent of the contents of this section.

By simple power-counting, the problem of a generalized
``Ginzburg-Landau'' theory in 3+1 dimensions (with a many-component
``superconducting'' field $\varphi_\ell$) is in its upper critical
dimension, so one expects either of two possibilities.  One
possibility is that the Gaussian fixed point is marginally stable, and
mean-field behavior is correct up to logarithmic factors.  The
other possibility is that the Gaussian fixed point is marginally
unstable, and the true critical behavior is a strong coupling problem;
most probably, such flows to strong coupling indicate a weak
fluctuation-induced first order transition.

Here we follow Balents {\sl et
  al.}\cite{Balents_1:prb05,Balents_2:prb05}, which generalized the
calculations of Halperin, Lubensky and Ma\cite{HLM:prl74} and Brezin {\sl et
al.}\cite{Brezin:prb74}, and consider a general $q$-component $[U(1)]$
action \be
\begin{split} &
{\mathcal S}_0 = \int d^D r 
\Big\{ \sum_{\ell=0}^{q-1}
|(\partial_{\mu} -i \alpha_{\mu}) \varphi_{\ell}|^2 + s |\varphi_{\ell}|^2
+ \frac{1}{2 e^2} F^2 
\Big\} \; , \\ &
{\mathcal S}_1 = \frac{1}{4} \int d^D r
\sum_{\ell,m,n,i=0}^{q-1}
u_{\ell m;n i}^{\phantom{*}} \varphi_{\ell}^* \varphi_m^* \varphi_n^{\phantom{*}} \varphi_i^{\phantom{*}}
\; .
\end{split}\label{eq:generalaction}
\ee Here we have written the theory for a general space-time
dimensionality $D$.  For the quantum critical point of interest,
$D=3+1=4$ total space-time dimensions.  For this very general action
the RG flows obtained by an $\epsilon$-expansion
are\cite{Balents_1:prb05,Balents_2:prb05} \be
\begin{split} &
\frac{de^2}{d\ell} = \epsilon e^2 - \frac{C q}{3} e^4 \; , \\ &
\frac{du_{\ell m;n k}}{d\ell} = 
(\epsilon + 2 \eta)u_{\ell m;n k} - 
a_2 e^4 C \left[ 
\delta_{\ell n} \delta_{m k} + \delta_{\ell k} \delta_{m n}
\right] \\ & -
C \sum_{ij} \Big[
\frac{1}{2} u_{\ell m;i j} u_{i j;n k} +
u_{\ell i;n j} u_{m j;k i} + u_{\ell i;k j} u_{m j;n i}
\Big] \; , \\ &
\eta = a_1 e^2 C
\; ,
\end{split}
\ee
with {\emph{spatial}} dimension $d = 3 - \epsilon$. Here $\eta = a_1 e^2 C$ is the anomalous dimension of the fields. The constants $C , a_{1,2}$ were calculated\cite{Balents_1:prb05,Balents_2:prb05} in an $\epsilon$-expansion. For our case $\epsilon=0$ since our model is in $3+1$ spacetime dimensions. Thus the $\epsilon$-expansion
results hold exactly:
$ C= \frac{1}{8\pi^2} , \quad a_1 =3 , a_2 = 6 $.

We take directly $\varphi \rightarrow \zeta , \xi$ in the action \eqref{action3} to make contact with the couplings $\gamma_j$
\be
\begin{split} &
u_{m m;m m} = 4 \gamma_1 \; , \\ &
u_{m n;m n} = 4 (\gamma_2 + \frac{1}{2} \gamma_1) \quad n \neq m \; , \\ &
u_{m \bar{n};m \bar{n}} = (\gamma_3 + 2 \gamma_1) \; , \\ &
u_{a \bar{b};c \bar{d}} = \gamma_4 \kappa_{a \bar{b}}^{\phantom{a \bar{b}} c \bar{d}}
\quad a \neq c , b \neq d
\; ,
\end{split}
\ee where the indices $\in 0...3$ and the overlined indices denote the
$\xi_j$ quartet of field components. All other $u_{ijkl}$ couplings
vanish in our special case. This ``structure'' encodes in it the
specific symmetry group of our lattice model. Therefore, the RG flow
equations should preserve this general structure -- only 4 independent
couplings, and the $\kappa$-tensor must retain its structure.  This
provides a good check that we have indeed included all possible quartic
terms allowed by the symmetry of the problem.
 
Some manipulation gives us the flow equations for our 4 couplings: 
\be
\begin{split}
\frac{d\gamma_1}{d\ell} = &
-C\, \big( 3\,e^4 - 6\,e^2\,{{\gamma }_1} + 8\,{{{\gamma }_1}}^2 + 
\\ &
      16\,{{\gamma }_1}\,{{\gamma }_2} + 16\,{{{\gamma }_2}}^2 + 
      8\,{{\gamma }_1}\,{{\gamma }_3} + 2\,{{{\gamma }_3}}^2 
\big)
\; ,   \\ 
\frac{d\gamma_2}{d\ell} = &
C\,\left( 6\,e^2\,{{\gamma }_2} - 2\,{{\gamma }_1}\,{{\gamma }_2} + 
    2\,{{{\gamma }_2}}^2 - {{{\gamma }_4}}^2 \right)
\; ,   \\ 
\frac{d\gamma_3}{d\ell} = &
2\,C\, \big( 16\,{{{\gamma }_2}}^2 + 3\,e^2\,{{\gamma }_3} - 
\\ &
    6\,{{\gamma }_1}\,{{\gamma }_3} - 8\,{{\gamma }_2}\,{{\gamma }_3} + 
    {{{\gamma }_3}}^2 - 3\,{{{\gamma }_4}}^2 \big)
\; ,    \\ 
\frac{d\gamma_4}{d\ell} = &
2\,C\,\left( 3\,e^2 - 6\,{{\gamma }_1} - 2\,{{\gamma }_3} - 
    {{\gamma }_4} \right) \,{{\gamma }_4}
\; .
\end{split}
\ee
The only fixed point allowed by the RG equations is a trivial fixed point 
with all coupling strengths vanishing $\gamma_i = 0, e=0$.

The stability or lack thereof of such coupled non-linear differential
equations is not obvious.  We would like to know if there is a
subspace of codimension zero of the four dimensional phase space (we
may project it onto the $e=0$ plane since the evolution $e>0$ is
clearly monotonically decreasing toward zero) in which all couplings
scale toward zero.  While we have not been able to prove this is not
the case, our numerical and analytical investigations have found no
such stable regime.  We {\sl did} find a few specific fine-tuned
stable solutions, but these were all unstable to infinitesimal
perturbations.  Thus we believe that the mean-field critical behavior
is always destabilized by fluctuations.  We expect this probably
signals a fluctuation-driven weakly first order transition.
Intuitively, this is a result of the gauge field
fluctuations~\cite{HLM:prl74}, which lead to attractive interactions
amongst the monopoles, driving bound state formation.  Note that this
result is reliable only for the zero temperature transition with
$D=3+1=4$, the upper critical dimension, where the perturbative
renormalization group treatment is justified.  In $D=3+0$, the same
field theory may well have non-trivial stable (if one parameter is
tuned to criticality) fixed points.  We will return to this point in
the discussion below.

\section{Discussion}
\label{sec:conclusions}

In the preceding sections, we have presented a systematic study of the
zero temperature ordered phases proximate to a ``half-polarized''
(from the 3:1 constraint) $U(1)$ spin liquid on the pyrochlore
lattice, based on a projective symmetry group (PSG) analysis of the
monopole excitations of the liquid state.  One of these states would
be expected to occur on reducing $v$ in the spin/dimer model starting
from a value ($\lesssim 1$) within the spin liquid state.  The ordered
phases, determined at the mean-field level, all break discrete
symmetries of the pyrochlore lattice, and in particular we find a
generic unit cell enlargement, with the minimal cell size 4 times that
of the underlying pyrochlore lattice, and a maximal unit cell 64 times
larger.  The simplest ``R'' state, with the smallest unit cell, is the
same one which was found to be the ground state in the ``classical''
spin/dimer limit $v\rightarrow -\infty$.  It therefore seems likely
that the spin/dimer model exhibits only two phases for $v<1$, and
hence a {\sl direct} quantum phase transition from the U(1) spin
liquid to the R state.  If this is indeed the case, one may
contemplate the possibility that \hgaf\ (or other pyrochlore
chromates) might be close to this quantum critical point.

\begin{figure}
        \centering
                \ifig[width=3.5in]{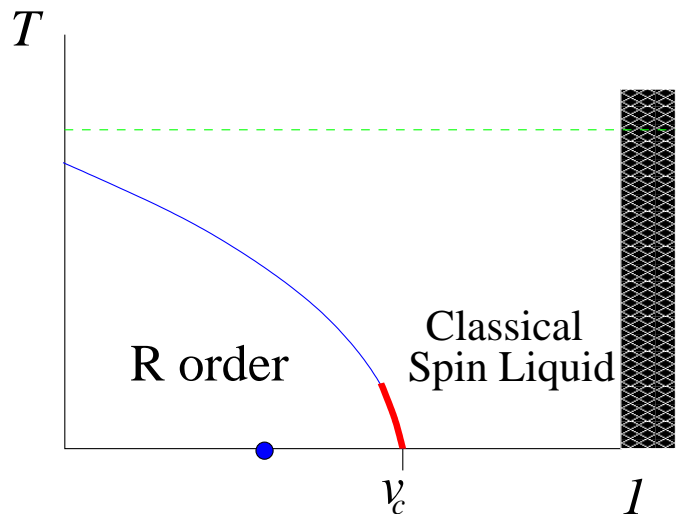}
        \caption{(Color online.) Schematic phase diagram of the
          spin-dimer model in the $v$-$T$ plane.  At zero temperature,
          we suppose there is a direct transition from a U(1) spin
          liquid to the R state at $v=v_c<1$.  The thick (red online)
          portion of the $T>0$ phase boundary is first order, while
          the thin (blue online) boundary denotes a possible
          non-trivial ``non-LGW'' second order transition at higher
          temperature.}
        \label{fig:thermalpd}
\end{figure}

While previous work\cite{Bergman:prl05} and the main text has focused
almost entirely upon the ground state properties of the spin/dimer
model, it is interesting to consider the more general problem at
non-zero temperature $T>0$.  A schematic phase diagram in the $v$-$T$
plane is illustrated in Fig.~\ref{fig:thermalpd}.  We will focus on
the region $v<1$, and will not discuss the physics of the thermal
ensemble of ``frozen'' states occuring for $v>1$.  First, we note that
the R state has a gap to all excitations, and breaks only discrete
lattice symmetries.  We therefore expect that the R state will persist
at $T>0$ up to some non-zero critical temperature $T_c(v)>0$ for
$v<v_c$.  By contrast, the U(1) quantum spin liquid ground state
breaks no symmetries.  Therefore, in the dimer model, no transition is
expected for $v_c<v<1$ as $T$ is increased from zero to infinity.  We
thus expect just the single phase boundary emanating from the quantum
critical point, shown in Fig.~\ref{fig:thermalpd}.

It might appear from these observations that the {\sl thermal} ordering
transition at $T_c(v)$ should be a rather ordinary one, described by
the usual Landau-Ginzburg-Wilson (LGW) approach based on the order parameter
of the R state.  In fact this is incorrect, and $T>0$ problem is
rather more interesting.  To see this, let us consider the
``paramagnetic'' state obtained for very high temperature in the dimer
model.  The physics of such {\sl classical, infinite temperature}
dimer models (and other similarly constrained models) have been
considered by several authors\cite{Hermele:prb04,Isakov,Huse}. As shown
in these works, due to the dimer constraint, even at infinite
temperature the dimer model has non-trivial power-law ``dipolar''
correlations.  Such dipolar correlations are not captured by the
conventional LGW theory which retains only the order parameter.  

To understand these dipolar correlations, {\sl and} the a proper
formulation of the phase transition in the spin/dimer model, it is
instructive to return to the monopole field theory in
Eq.(\ref{eq:generalaction}). Let us rewrite this effective action
making space and time coordinates separate and explicit:
\begin{eqnarray}
  \label{eq:genact2}
&& {\mathcal S}_0  =  \nonumber \\  && \int d^{3}r\!\! \int_0^\beta d\tau \!\!
\Big\{ \sum_{\ell=0}^{q-1}
|(\partial_\tau -i \alpha_0) \varphi_{\ell}|^2 + |(\vec\nabla -i
\vec\alpha) \varphi_{\ell}|^2 + s |\varphi_{\ell}|^2 \nonumber \\
& & + \frac{1}{2 e^2} \left( \partial_\tau \vec{\alpha} - \vec\nabla
  \alpha_0\right)^2 + \frac{1}{2e^2}\left( \vec\nabla \times
  \vec\alpha\right)^2 
\Big\} \; , \nonumber \\ 
&& {\mathcal S}_1  =  \frac{1}{4} \int d^{3}r \! \int_0^\beta  d\tau \!
\sum_{\ell,m,n,i=0}^{q-1}
u_{\ell m;n i}^{\phantom{*}} \varphi_{\ell}^* \varphi_m^* \varphi_n^{\phantom{*}} \varphi_i^{\phantom{*}}
\; .
\end{eqnarray}
Here $\beta = (k_{\scriptscriptstyle B}T)^{-1}$ is the inverse temperature.
To derive a theory of the thermal phase transition at $T>0$, we use
standard logic to proceed from Eq.~(\ref{eq:genact2}).  In particular,
at $T>0$, imposing periodic boundary conditions (as demanded by the
trace defining the quantum statistical mechanical partition function)
in imaginary time leads to a set of discrete bosonic Matsubara
frequencies $\omega_n = 2\pi n/\beta$, with integer $n$.  Because of
the time-derivative term, all modes with $\omega_n \neq 0$ have
enhanced ``masses'' relative to the zero Matsubara frequency mode, and
can be integrated out.  Practically speaking, this amounts to assuming
the order parameter is constant in imaginary time, $\partial_\tau
\varphi_\ell=0$.  Similarly, we may take $\partial_\tau \vec\alpha=0$,
and by a choice of gauge, $\alpha_0=0$.  Carrying out this procedure,
we find $\mathcal{S}_0+\mathcal{S}_1 =
\mathcal{F}/k_{\scriptscriptstyle B}T$, where
$\mathcal{F}=\mathcal{F}_0+\mathcal{F}_1$ is an effective classical
free energy: 

\begin{eqnarray}\label{eq:free1}
 {\mathcal F}_0  & = & \int d^{3}r
\Big\{ \sum_{\ell=0}^{q-1}
|(\vec\nabla -i
\vec\alpha) \varphi_{\ell}|^2 + s |\varphi_{\ell}|^2 \nonumber \\
& & + \frac{1}{2e^2}\left( \vec\nabla \times
  \vec\alpha\right)^2 \Big\} \; , \nonumber \\ 
 {\mathcal F}_1  & = &  \frac{1}{4} \int d^{3}r \! 
\sum_{\ell,m,n,i=0}^{q-1}
u_{\ell m;n i}^{\phantom{*}} \varphi_{\ell}^* \varphi_m^* \varphi_n^{\phantom{*}} \varphi_i^{\phantom{*}}
\; .%\nonumber
\end{eqnarray}

Eq.~(\ref{eq:free1}) is precisely the classical ``Ginzburg-Landau''
free energy for a multi-component superconductor in three dimensions,
with the quartic interaction $u_{\ell m;n i}$ determined by the PSG.
The interpretation of this result is quite simple.  In the U(1) spin
liquid, the monopole is a well-defined, bosonic particle excitation,
and carries the ``magnetic'' gauge charge.  The magnetic gauge charge
is conserved in the theory.  As already discussed, the ordered (R)
state at zero temperature is understood as a condensate of these
monopoles.  In fact, within the quantum dimer model, the bosonic
monopole can condense at a non-zero temperature, just as in an
ordinary Bose-Einstein condensate.  Thus Eq.~(\ref{eq:free1}) is
nothing but the ``classical'' free energy describing the
``superfluid'' transition of this monopole.  Because it carries a
non-zero magnetic gauge charge, it is coupled to the electric vector
potential $\vec{\alpha}$, just as the electric charged Cooper pair
condensate is coupled to the magnetic vector potential in ordinary
superconductivity.

Thus we are led to this remarkable and unconventional description of
the $T>0$ phase transition between the paramagnet and the R state.
Without the 3:1 spin/dimer constraint, this transition would certainly
be expected to be governed by conventional Landau theory.  What is the
nature of the unconventional transition?  According to the
$d=4-\epsilon$ expansion approach (discussed in
Sec.~\ref{sec:rg-analysis}), this transition is fluctuation-driven
first order.  However, this conclusion is known to be often incorrect
for the physical case of Ginzburg-Landau transitions in three
dimensions.  In particular, a class of related models, $N$-component
Ginzburg-Landau theories with $U(N)$ symmetry, have been investigated
in a number of cases.  First, for sufficiently large $N$, these
transitions can be shown to be second order in an expansion around
$N=\infty$.  Second, for $N=1$, a duality transformation has been used
to demonstrate that the transition can be continuous, in the inverted
XY universality class\cite{DasguptaHalperin}. A similar duality
analysis, in conjunction with numerics, has been used in arguing for
continuous critical behavior for $N=2$\cite{MotrunichVishwanath}. It therefore
appears quite likely that continuous critical behavior is possible in
these models for {\sl any} $N$.  By analogy, continuous critical
behavior of the theory of Eq.~(\ref{eq:free1}) seems quite possible.
We thus suggest that the paramagnetic to R state transition in the
spin/dimer model constitutes a novel non-LGW universality class.  Note
that, since the RG analysis in Sec.~\ref{sec:rg-analysis} concluded
that the $T=0$ quantum phase transition is weakly first order, the
phase boundary for {\sl very} small but non-zero temperature must
remain discontinuous.  We therefore expect a multicritical point
separating on the $T>0$ phase boundary separating first order from
continuous non-LGW critical behavior.  The continuous and first order
portions of the phase boundary are shown in Fig.~\ref{fig:thermalpd}
by thin (blue online) and thick (red online) lines, respectively.  It
would be of considerable interest to investigate this classical phase
transition in the the spin/dimer model numerically.  This could be
done on a purely classical dimer model, with $K=0$, and so would
require only classical Monte Carlo methods.

As remarked above, {\sl without} the 3:1 spin/dimer constraint, the
classical phase transition would certainly be expected to be described
by LGW theory.  For any microscopic model, such as the spin-$3/2$
Heisenberg model of Ref.\onlinecite{Bergman:prl05}, the 3:1 constraint is
not expected to be exactly obeyed.  At zero temperature, however,
pyrochlore tetrahedra violating the 3:1 constraint are gapped
excitations, and are not present in the ground state on the plateau.
Consider configurations in which a single tetrahedron violates the
constraint by having either zero or two minority spins instead of one.
In the mapping to the gauge theory, these particular excitations can
be viewed as states with ``electric'' gauge charge $\pm 1$ (relative
to the static background gauge charge of the plateau states) on the
tetrahedron in question.  They also carry physical spin $\Delta
S^z=\pm 3/2$ relative to the plateau states.  This follows because the
total spin can be written as
\begin{equation}
  \label{eq:Stot}
  S^z_{\rm TOT} =\frac{1}{2} \sum_t S_t^z,
\end{equation}
the factor of $1/2$ being required since each spin is contained in two
tetrahedra.  These excited states can therefore be viewed as
fractional spin excitations, or ``spinons'', in this particular
example.  In any case, because they cost only finite energy, there
will be a non-vanishing concentration of such electric gauge charges
at $T>0$, due to thermal activation.  The typical separation of the
electric gauge charges is expected to behave exponentially at low
temperature, $\lambda \sim \exp (\Delta/k_{\scriptscriptstyle B}T)$,
if $\Delta$ is the gap to the electric charged particles.  This has
several effects.  First, the dipolar correlations of the paramagnetic
phase of the spin/dimer model will cross over to the usual exponential
ones of an ordinary paramagnet, for lengths larger than $\lambda$.
Moreover, the ``plasma'' of electric charges is expected to give rise
to a linear {\sl confining} potential between oppositely
(magnetically) charged monopoles.  This will bind the monopoles and
anti-monopoles into gauge-neutral pairs, the radius of this bound
state being at least as large as $\lambda$.  It is only these pairs
that can Bose condense.  When $\lambda$ is large, there is a
cross-over behavior.  On approaching the paramagnetic to R
state transition, the correlation length grows in a manner first
governed by the non-LGW theory of Eq.~(\ref{eq:free1}).  Once the
correlation length exceeds the monopole-antimonopole binding
length, these pairs may be considered well-formed, and the critical
behavior changes to that described by Bose condensation of the pairs.
The creation/annihilation operators for the monopole-antimonopole
bound states is expected to be proportional to
\begin{equation}
  \label{eq:pairfield}
  \Psi_{\ell\ell'} \sim \varphi_\ell^* \varphi_{\ell'}^{\vphantom*}.
\end{equation}
Since $\varphi_\ell^*$ and $\varphi_\ell$ transform under conjugate
representations of the PSG, and hence $\Psi_{\ell\ell'}$ is
gauge-neutral, it transforms not under the PSG but simply the ordinary
lattice space group.  Hence $\Psi_{\ell\ell'}$ can be decomposed into
irreducible representations of the space group, which are precisely
the usual Landau order parameters!  Thus the critical behavior
sufficiently close to the phase boundary, when the 3:1 spin/dimer
constraint is not rigidly enforced, is indeed governed by LGW theory
as expected on general grounds.  The conventional LGW analysis is
straightforward, and will be presented in Ref.~\onlinecite{longpaper}.
It predicts that the paramagnetic to R state transition should be
first order, due to the presence of a cubic invariant.  For
$k_{\scriptscriptstyle B}T \ll \Delta$, it will be weakly so along the
thin (blue) portion of the phase boundary in Fig.~\ref{fig:thermalpd}.
Of course there is no such crossover in the spin/dimer model, for
which the constraint is rigidly enforced.
  
In summary, we have studied the phase structure of a spin/dimer model
on the pyrochlore/diamond lattice.  It contains an interesting quantum
paramagnetic ``U(1) spin liquid'' phase at zero temperature, as well
as an ordered ``R'' state at $T\geq 0$. We derived a novel monopole
field theory to describe the quantum and classical phase transitions
between the quantum and classical paramagnets and the R state.  Prior
work in Ref.\onlinecite{Bergman:prl05} indicates that the half-polarized
magnetization plateau of the spin-$3/2$ Heisenberg antiferromagnet on
the pyrochlore lattice is described by this model, with a coupling
constant that places it in the R state at low temperature.  It is
interesting to contemplate the possibility that the experimental
materials \cdaf\,, \hgaf\, might be near the quantum phase transition of
the paper to the spin liquid state.  It would indeed be exciting were
some homologous material to actually realize the U(1) spin liquid in
its plateau ground state.  To this end, we note that, were a
spin-$1/2$ pyrochlore antiferromagnet to be realized experimentally,
the quantum effects would be significantly further increased, and a
U(1) spin liquid might well be expected theoretically.  We leave such
tantalizing possibilities as open questions.

\acknowledgments

We are grateful to A. A. Burkov, M. Hermele, O. I. Motrunich, and
R. Shindou for enlightening discussions. This work was generously
supported by NSF DMR04-57440, NSF PHY99-07949, and the Packard
Foundation.

\appendix

\section{Ground state eigenmodes }
\label{app:GS_modes}

The diamond lattice is bipartite and can be thought of as being made
of 2 FCC sub-lattices one shifted from the other by the vector $ {\vec
  b} = -\frac{1}{4} ( {\vec a}_1 + {\vec a}_2 + {\vec a}_3 ) $ where
${\vec a}_j$ are the FCC primitive vectors. We introduce the notation
\be
\begin{split} &
\eta_1 (\vec n)= \psi ({\vec r})\;,  \qquad 
\\ &
\eta_2 (\vec n)= \psi ({\vec r} + {\vec b})
\; ,
\end{split}
\ee
where $\vec r$ denotes sites on an FCC lattice, $\vec r + \vec b$ are the sites of the second FCC lattice
comprising the diamond lattice, and $\vec n$ are the ``index'' space coordinates on the FCC lattice.

Defining the Fourier transform of the wavefunction $ \eta $, and its 
inverse as
\be \label{Fourier_defined}
\begin{split} &
{\eta}(\vec n) = \int\limits_{{\vec k} \in BZ} \! \! \! {\frac {d^3 {\vec k}} {(2\pi)^3}}
{\eta}(\vec k) \cdot  
e^{+i {\vec k} \cdot {\vec n}}
\; ,
\\ &
{\eta}(\vec k) = 
%{\frac{1}{\sqrt N}}
\sum_{\vec n} 
e^{-i {\vec k} \cdot {\vec n}}
\cdot {\eta}(\vec n)
\; ,
\end{split}
\ee
we can write out the eigenstates in a compact manner
\be \label{gs_nu}
\begin{split}
\Phi_{\nu} = 
\sum_{\substack {{\mu=0...3} \\ {a=1,2}}}
c^{(a)}_{\mu} \, \eta_{a} ({\vec p}_{\mu ,\nu})
\, e^{-i \frac{\pi}{2} \mu \nu}
\; ,
\\
{\overline \Phi}_{\nu} = 
\sum_{\substack {{\mu=0...3} \\ {a=1,2}}}
c^{(\bar a)}_{- \mu} \, \eta_{a} ({\vec q}_{\mu ,\nu})
e^{-i \frac{\pi}{2} \mu \nu}
\; ,
\end{split}
\ee
where $\ket 0 $ denotes the vacuum state, and
\be
\begin{split}
{\vec p}_{\mu ,\nu} = {\vec p} + \mu {\vec Q} + \frac{\pi}{2} \nu {\vec \epsilon}
\; ,
\\
{\vec q}_{\mu ,\nu} = {\vec d} - {\vec p} + \mu {\vec Q} + \frac{\pi}{2} \nu {\vec \epsilon}
\; ,
\end{split}
\ee
where
$ {\vec p} = ( \frac{\pi }{2},\frac{\pi }{4},\frac{\pi }{4} ) $ and
$ {\vec d} = ( \frac{\pi }{2},0,\pi) $ .
The coefficients are
\be \label{relations}
c^{(a)}_{\mu,\nu} = 
c^{(a)}_{\mu}
\, e^{-i \frac{\pi}{2} \mu \nu}
\; ,
\ee
where $c^{(a)}_{\mu} = c^{(a)}_{\mu,0}$ are the coefficients for the $\nu = 0$ state.
Finally, the un-normalized coefficients $ c_{\mu}^{(a)} $ are:
\be \label{c_mu_a}
\bs &
c_{\mu}^{(1)} = 
\left(
1, -i, \frac{-i}{1+{\sqrt 2}} , \frac{1}{1+{\sqrt 2}} 
\right)
\; ,
\\ &
c_{\mu}^{(2)} = 
\left(
\frac{3-i}{\sqrt{10}} , \frac{1-2i}{\sqrt 5 + \sqrt{10}}, - \frac{1+3i}{{\sqrt{10}} (1+{\sqrt 2})} , \frac{2+i}{\sqrt 5} 
\right)
\; .
\end{split}
\ee

\section{Projective symmetry group (PSG) }
\label{app:PSG}

We start by enumerating some diamond lattice symmetries.
Since the diamond lattice is made up of two FCC lattices, it inherits the translations of the FCC lattice
\be
{\hat {\mathit t}}_j : \, \vec r \longrightarrow   
\vec r + {\vec a}_j
\; .
\ee
Next we consider a 3-fold rotation symmetry
\be
{\mathit r}_3 : 
(x,y,z) \longrightarrow  (z,x,y)
\; ,
\ee
and a 2-fold rotation
\be
{\mathit r}_2 : (x,y,z) \longrightarrow  (x,-y,-z) 
\; .
\ee
There is also a reflection symmetry
\be
{\mathit i} : \left( x,y,z \right) \longrightarrow
\left( x,z,y \right)
\; .
\ee

Finally we have an inversion symmetry that effectively swaps between the two FCC sublattices
\be
{\mathit p} : 
\vec r 
\longrightarrow \vec b - {\vec r}
\; .
\ee

To construct the PSG we attach gauge transformations to each symmetry operation. This is most easily done
in the ``index'' space coordinates, and we now list the action of the PSG operations on the
wavefunctions.
First are the three lattice translations
\be
\T{j} : \, {\eta}(\vec n) \longrightarrow   
{\eta}(\vec n + {\vec a}_j)
e^{-i \Lambda_j ({\vec n}) }
\; ,
\ee
where
\be
 \Lambda_j ({\vec n}) = - \left( {\vec \epsilon} \cdot {\vec n} \right) Q_j 
 \; .
\ee 
The inversion symmetry
\be
\C : \, {\eta}(\vec n) \longrightarrow   
{\hat \sigma}_x \cdot {\eta}(-{\vec n})
e^{-i \Lambda_{\mathcal C} ({\vec n}) }
\; ,
\ee
where $\sigma_x$ is the $x$ Pauli matrix, and
\be \label{C_gauge}
\Lambda_{\mathcal C} ({\vec n}) = 
- {\vec d} \cdot {\vec n}
= 
- {\frac{\pi}{2}} 
\left( 1,0,-2 \right) \cdot {\vec n}
\; .
\ee
The 3-fold rotation
\be
\P : \, {\eta}(\vec n) \longrightarrow   
\eta (\P \cdot {\vec n})
e^{-i \Lambda_{{\mathcal R}_{3}} (\vec n)}
\; ,
\ee
with the gauge transformation
\be
\Lambda_{{\mathcal R}_{3}} ({\vec n}) =  {\vec n} \cdot {\hat B_1} \cdot {\vec n} + 
{\vec \delta} \cdot {\vec n}
\; ,
\ee 
where
\be
{\hat B_1} = \frac {\pi} {4},
\begin{pmatrix}
1 & 1 & 2 \cr
1 & 2 & 1 \cr
2 & 1 & 1 \cr
\end{pmatrix}
\qquad
{\vec \delta} = - \frac{\pi}{4}
\left( 1,2,1 \right)
\; .
\ee
The 2-fold rotation
\be
\I : 
\Biggl\{
\begin{array}{ll}
\, {\eta_{1}}(\vec n) \longrightarrow   
{\eta_{1}}(\I \cdot {\vec n})
e^{-i \Lambda^{(1)}_{{\mathcal R}_{2}} ({\vec n}) }
\\
{\eta_{2}}(\vec n) \longrightarrow   
{\eta_{2}}(\I \cdot {\vec n} + {\vec a}_{1})
e^{-i \Lambda^{(2)}_{{\mathcal R}_{2}} ({\vec n}) }
\; .
\end{array}
\ee
The gauge transformations for the 2 spinor components are different
\begin{subequations} \label{R2gauge}
\begin{align}
& 
\Lambda^{(1)}_{{\mathcal R}_{2}} ({\vec n}) =  
{\vec n} \cdot {\hat B_2} \cdot {\vec n} + 
{\vec \lambda_1} \cdot {\vec n} 
\; ,
\\
&
\Lambda^{(2)}_{{\mathcal R}_{2}} ({\vec n}) =  
{\vec n} \cdot {\hat B_2} \cdot {\vec n} + 
{\vec \lambda_2} \cdot {\vec n}
\; ,
\end{align}
\end{subequations} 
where
\be
{\hat B_2} = \frac{\pi}{2}
\begin{pmatrix}
 0 & 1 & 1 \cr 1 & 1 & 0 \cr 1 & 0 & 1 \cr
\end{pmatrix}
\; ,
\ee
and
\be
{\vec \lambda_1}=\frac{\pi}{2}
\left( -1,1,2 \right)
\qquad
{\vec \lambda_2}={\vec \lambda_1}+\frac{\pi }{2}
\left( 1,2,1 \right)
\; .
\ee
Finally, the reflection symmetry
\be
\R : \, {\eta}(\vec n) \longrightarrow   
{\eta}^{\dag} (\R \cdot {\vec n}) \,
e^{-i \Lambda_{\R} \left( \vec n \right)}
\; ,
\ee
and the gauge transformation is
\be
\Lambda_{\R} ({\vec n}) =  {\vec n} \cdot {\hat B_3} \cdot {\vec n} + 
{\vec \xi} \cdot {\vec n}
\; ,
\ee 
where
\be
{\hat B_3} = - \frac {\pi} {4}
\begin{pmatrix}
2 & 1 & 1 \cr
1 & 2 & 1 \cr
1 & 1 & 2 \cr
\end{pmatrix},
\qquad
{\vec \xi} = \frac{\pi}{2}
\left( 1,1,1 \right)
\; .
\ee

One can take the Fourier transform of these symmetry operations, to
find their action on ${\eta}(\vec k)$ and then deduce how the 8 ground
state eigenmodes transform under them. The transformation rules
\eqref{eq:prep}-\eqref{eq:prep_f} are the result of that analysis. It is a
good check that the ground state manifold is invariant under every one
of these symmetries. Although the ground state manifold need not be
completely connected by the symmetry operations, in our case it is.

\section{Group algebra}
\label{app:algebra}

The diamond lattice symmetry group, described by 
the reduced set of operations introduced in the text,
obeys the algebra
\be\label{lattice_algebra}
\begin{split}
{{\mathit r}_3}^3 & = 1
\\
{{\mathit r}_2}^2 & = 1
\\
{{\mathit p}}^2 & = 1
\\
{{\mathit i}}^2 & = 1
\\
{\mathit r}_3^{\phantom{2}} \cdot {\mathit p}^{\phantom{2}} \cdot {{\mathit r}_3}^{-1} \cdot {\mathit p}^{\phantom{2}}  & = 1
\\
{\mathit t}_i^{\phantom{2}} \cdot {\mathit t}_j^{\phantom{2}} \cdot {{\mathit t}_i}^{-1} \cdot {{\mathit t}_j}^{-1} & = 1
\\
\left( {\mathit p} \cdot {\mathit t}_j \right)^{2} & = 1
\\
{\mathit r}_3^{\phantom{2}} \cdot {\mathit t}_j^{\phantom{2}} \cdot {{\mathit r}_3}^{-1} \cdot {{\mathit t}_{j+1}}^{-1} & = 1
\\
{\mathit r}_2 \cdot {\mathit t}_1 \cdot {\mathit r}_2 \cdot {{\mathit t}_1} & = 1
\\
{\mathit r}_2 \cdot {\mathit t}_2 \cdot {\mathit r}_2 \cdot {\mathit t}_1 & = {\mathit t}_3
\\
{\mathit r}_2 \cdot {\mathit t}_3 \cdot {\mathit r}_2 \cdot {\mathit t}_1 & = {\mathit t}_2
\\
{\mathit i} \cdot {\mathit t}_1 & = {\mathit t}_1 \cdot {\mathit i}
\\
{\mathit i} \cdot {\mathit t}_2 & = {\mathit t}_3 \cdot {\mathit i}
\\
{\mathit i} \cdot {\mathit t}_3 & = {\mathit t}_2 \cdot {\mathit i}
\\
{\mathit i}^{\phantom{2}} \cdot {\mathit r}_3^{\phantom{2}} & = {{\mathit r}_3}^{-1} \cdot {\mathit i}^{\phantom{2}}
\\
{\mathit i} \cdot {\mathit r}_2 & = {\mathit r}_2 \cdot {\mathit i}
\\
{\mathit i} \cdot {\mathit p} & = {\mathit p} \cdot {\mathit i}
\end{split}
\ee

\section{Ground state manifold permutative representation of PSG}
\label{app:irrep}

In the basis of \eqref{field_trans} the PSG realizes a permutative
representation,\cite{Balents_1:prb05} i.e. each generator can be
written as $G = \Lambda(G) P(G)$, where $\Lambda(G)$ is a diagonal
matrix with unimodular complex entries, and $P(G)$ is a permutation
matrix, acting on the 8-component vector $(\vec{\zeta},\vec{\xi})$.
One finds:
\begin{eqnarray}
  \label{eq:prep}
  \Lambda(\T{1}) & = & {\rm diag}(i,i,1,-1,1,1,-i,i), \nonumber \\
  P(\T{1}) & = & (43218765), \\
  \Lambda(\T{2}) & = & {\rm diag}(z,z,iz,iz,-iz,-iz,z,z), \nonumber \\
  P(\T{2}) & = & (21436587), \\
  \Lambda(\I) & = & {\rm diag}(z^*,z,-z^*,-z,z,z^*,z^*,z), \nonumber \\
  P(\I) & = & (21437856), \\
  \Lambda(\P) & = & {\rm diag}(z,1,-z^*,-1,z,1,z^*,1), \nonumber \\
  P(\P) & = & (32417685), \\
  \Lambda(\C) & = & {\rm diag}(1,1,1,-1,1,1,1,-1), \nonumber \\
  P(\C) & = & (56781234), \\
  \Lambda(\R) & = & {\rm diag}(w,w,iw,-w,-w,w,iw,w), \nonumber \\
\label{eq:prep_f}
  P(\R) & = & (86754231).
\end{eqnarray}
Here $z=e^{i\pi/4},w=(1-2i)/\sqrt{5}$, ${\rm diag}(\cdot)$ is the
diagonal matrix with entries $\cdot$, and permutations are specified
in the standard way by the result of permuting the integers $1\ldots
8$.  Finally, $\R$ should be understood as acting after complex
conjugation of the monopole field vector
\be
\R : (\vec{\zeta},\vec{\xi}) \rightarrow
\Lambda(\R) P(\R) ({\vec{\zeta}}^*,{\vec{\xi}}^*).
\ee

\section{$\kappa$-tensor}
\label{app:kappa_tensor}

The $\kappa$ [in Eq.~\eqref{eq:kappa_text}] tensor was hiding the following  
expression

\be
\begin{split}
\Theta_4 = \sum_{ijkl} \kappa_{i j}^{\phantom{i j} k l} \zeta_k^{\phantom{*}} \xi_l^{\phantom{*}} \zeta_{i}^* \xi_{j}^* 
\qquad \qquad \qquad \qquad \qquad \qquad \qquad \qquad 
\\ =  
{{\zeta}_1}\,{{\xi}_1}\,{{{{\zeta}_0}}^*}\,{{{{\xi}_0}}^*} + 
  {{\zeta}_2}\,{{\xi}_2}\,{{{{\zeta}_0}}^*}\,{{{{\xi}_0}}^*} + 
  i \,{{\zeta}_3}\,{{\xi}_3}\,{{{{\zeta}_0}}^*}\,{{{{\xi}_0}}^*} - 
  {{\zeta}_0}\,{{\xi}_1}\,{{{{\zeta}_1}}^*}\,{{{{\xi}_0}}^*} \\ + 
  i \,{{\zeta}_3}\,{{\xi}_2}\,{{{{\zeta}_1}}^*}\,{{{{\xi}_0}}^*} + 
  {{\zeta}_2}\,{{\xi}_3}\,{{{{\zeta}_1}}^*}\,{{{{\xi}_0}}^*} - 
  {{\zeta}_3}\,{{\xi}_1}\,{{{{\zeta}_2}}^*}\,{{{{\xi}_0}}^*} + 
  {{\zeta}_0}\,{{\xi}_2}\,{{{{\zeta}_2}}^*}\,{{{{\xi}_0}}^*} \\ + 
  i \,{{\zeta}_1}\,{{\xi}_3}\,{{{{\zeta}_2}}^*}\,{{{{\xi}_0}}^*} - 
  {{\zeta}_2}\,{{\xi}_1}\,{{{{\zeta}_3}}^*}\,{{{{\xi}_0}}^*} + 
  i \,{{\zeta}_1}\,{{\xi}_2}\,{{{{\zeta}_3}}^*}\,{{{{\xi}_0}}^*} - 
  {{\zeta}_0}\,{{\xi}_3}\,{{{{\zeta}_3}}^*}\,{{{{\xi}_0}}^*} \\ - 
  {{\zeta}_1}\,{{\xi}_0}\,{{{{\zeta}_0}}^*}\,{{{{\xi}_1}}^*} - 
  {{\zeta}_3}\,{{\xi}_2}\,{{{{\zeta}_0}}^*}\,{{{{\xi}_1}}^*} - 
  i \,{{\zeta}_2}\,{{\xi}_3}\,{{{{\zeta}_0}}^*}\,{{{{\xi}_1}}^*} + 
  {{\zeta}_0}\,{{\xi}_0}\,{{{{\zeta}_1}}^*}\,{{{{\xi}_1}}^*} \\ - 
  i \,{{\zeta}_2}\,{{\xi}_2}\,{{{{\zeta}_1}}^*}\,{{{{\xi}_1}}^*} - 
  {{\zeta}_3}\,{{\xi}_3}\,{{{{\zeta}_1}}^*}\,{{{{\xi}_1}}^*} - 
  {{\zeta}_3}\,{{\xi}_0}\,{{{{\zeta}_2}}^*}\,{{{{\xi}_1}}^*} - 
  {{\zeta}_1}\,{{\xi}_2}\,{{{{\zeta}_2}}^*}\,{{{{\xi}_1}}^*} \\+ 
  i \,{{\zeta}_0}\,{{\xi}_3}\,{{{{\zeta}_2}}^*}\,{{{{\xi}_1}}^*} - 
  {{\zeta}_2}\,{{\xi}_0}\,{{{{\zeta}_3}}^*}\,{{{{\xi}_1}}^*} + 
  i \,{{\zeta}_0}\,{{\xi}_2}\,{{{{\zeta}_3}}^*}\,{{{{\xi}_1}}^*} + 
  {{\zeta}_1}\,{{\xi}_3}\,{{{{\zeta}_3}}^*}\,{{{{\xi}_1}}^*} \\+ 
  {{\zeta}_2}\,{{\xi}_0}\,{{{{\zeta}_0}}^*}\,{{{{\xi}_2}}^*} - 
  i \,{{\zeta}_3}\,{{\xi}_1}\,{{{{\zeta}_0}}^*}\,{{{{\xi}_2}}^*} + 
  {{\zeta}_1}\,{{\xi}_3}\,{{{{\zeta}_0}}^*}\,{{{{\xi}_2}}^*} - 
  i \,{{\zeta}_3}\,{{\xi}_0}\,{{{{\zeta}_1}}^*}\,{{{{\xi}_2}}^*}\\ - 
  {{\zeta}_2}\,{{\xi}_1}\,{{{{\zeta}_1}}^*}\,{{{{\xi}_2}}^*} + 
  {{\zeta}_0}\,{{\xi}_3}\,{{{{\zeta}_1}}^*}\,{{{{\xi}_2}}^*} + 
  {{\zeta}_0}\,{{\xi}_0}\,{{{{\zeta}_2}}^*}\,{{{{\xi}_2}}^*} + 
  i \,{{\zeta}_1}\,{{\xi}_1}\,{{{{\zeta}_2}}^*}\,{{{{\xi}_2}}^*}\\ + 
  {{\zeta}_3}\,{{\xi}_3}\,{{{{\zeta}_2}}^*}\,{{{{\xi}_2}}^*} - 
  i \,{{\zeta}_1}\,{{\xi}_0}\,{{{{\zeta}_3}}^*}\,{{{{\xi}_2}}^*} - 
  {{\zeta}_0}\,{{\xi}_1}\,{{{{\zeta}_3}}^*}\,{{{{\xi}_2}}^*} - 
  {{\zeta}_2}\,{{\xi}_3}\,{{{{\zeta}_3}}^*}\,{{{{\xi}_2}}^*} \\- 
  {{\zeta}_3}\,{{\xi}_0}\,{{{{\zeta}_0}}^*}\,{{{{\xi}_3}}^*} - 
  i \,{{\zeta}_2}\,{{\xi}_1}\,{{{{\zeta}_0}}^*}\,{{{{\xi}_3}}^*} + 
  {{\zeta}_1}\,{{\xi}_2}\,{{{{\zeta}_0}}^*}\,{{{{\xi}_3}}^*} - 
  i \,{{\zeta}_2}\,{{\xi}_0}\,{{{{\zeta}_1}}^*}\,{{{{\xi}_3}}^*} \\+ 
  {{\zeta}_3}\,{{\xi}_1}\,{{{{\zeta}_1}}^*}\,{{{{\xi}_3}}^*} + 
  {{\zeta}_0}\,{{\xi}_2}\,{{{{\zeta}_1}}^*}\,{{{{\xi}_3}}^*} + 
  {{\zeta}_1}\,{{\xi}_0}\,{{{{\zeta}_2}}^*}\,{{{{\xi}_3}}^*} + 
  i \,{{\zeta}_0}\,{{\xi}_1}\,{{{{\zeta}_2}}^*}\,{{{{\xi}_3}}^*} \\- 
  {{\zeta}_3}\,{{\xi}_2}\,{{{{\zeta}_2}}^*}\,{{{{\xi}_3}}^*} - 
  i \,{{\zeta}_0}\,{{\xi}_0}\,{{{{\zeta}_3}}^*}\,{{{{\xi}_3}}^*} - 
  {{\zeta}_1}\,{{\xi}_1}\,{{{{\zeta}_3}}^*}\,{{{{\xi}_3}}^*} + 
  {{\zeta}_2}\,{{\xi}_2}\,{{{{\zeta}_3}}^*}\,{{{{\xi}_3}}^*}
  \; .
\end{split}
\ee

\section{Proof that the lowest dimension of a representation of the PSG is 8}
\label{app:8dim_proof}

We start by assuming that we are in a basis where $\T{1}$ is diagonal. 
Denote one eigenvector by
\be
\T{1}\psi_0 = \lambda \psi_0 .
\ee
The PSG 
algebra rule \eqref{lattice_algebra3} dictates that $\T{2}$ cannot be diagonal in this basis.
Using 
\be
\T{2} \cdot \T{1} = {\T{1}} \cdot {\T{2}} (-i) \; ,
\ee
we find $\psi_0$ is connected to a 4--cycle of eigenvectors
\be
\psi_m = \T{2}^m \psi_0,
\ee
with eigenvalues
\be
\T{1}\psi_m = \lambda i^m \psi_m 
\; .
\ee
This proves that the translations must all be constructed only of 4--cycles. As a consequence, 
any representation of the PSG can only be of a dimension $d = 4 \times n$.

We know already that an 8 dimensional representation (rep) exists \eqref{eq:prep}-\eqref{eq:prep_f}, so 
we need only consider the $d=4$ case, and show that it is impossible.

Assume we have a $d=4$ rep. Therefore, each eigenvalue is non--degenerate, with a unique eigenvector. From 
\eqref{lattice_algebra} we know that the inversion and $\T{1}$ commute. It is easy to 
show that
\be
(\R \psi_m) \lambda i^m = \T{1} (\R \psi_m)
\ee
follows, and since only a unique eigenvector has this eigenvalue, we conclude that
\be
(\R \psi_m) = \psi_m
\ee
and that in this rep $\R = 1$. From the PSG algebra we now have
\be
\R \P = \P^{-1} \R \Rightarrow \P = \P^{-1}
\; .
\ee
The rotation $\P$ is 3--fold, and therefore in this rep it is also unity!
Finally, as $\P=1$ using the rule
\be
\T{2} = \P \T{1} \P^{-1}
\ee
we find $\T{1} = \T{2}$, and we have a contradiction.

This contradiction proves a 4--dim rep of the PSG is \emph{not} possible, leaving us with the $d=8$
as the lowest dimension of a rep of the PSG.

\end{document}